\documentclass[aps,twocolumn,showpacs,amsmath,amssymb,pre, longbibliography
]{revtex4-1}

\pdfoutput=1
\usepackage[utf8]{inputenc}
\usepackage{lmodern}		
\usepackage{cmap}			

\usepackage{lipsum}
\usepackage{calc}
\usepackage{xifthen}
\usepackage[]{algorithm2e}
\usepackage[table,xcdraw]{xcolor}
\usepackage{xr}
\usepackage{amssymb}
\usepackage{amsmath}
\usepackage{multirow}
\usepackage{todonotes}
\usepackage{graphics,graphicx}
\usepackage{xspace}
\usepackage{dsfont}
\usepackage{bm}
\usepackage{mathtools}
\usepackage{hyperref}
\usepackage{color, soul}
\newcommand{\figref}[1]{~Fig.\ref{#1}}
\newcommand{\eqreff}[1]{Eq.\eqref{#1}}

\hypersetup{
	colorlinks,%
	citecolor=blue,%
	filecolor=blue,%
	linkcolor=blue,%
	urlcolor=blue
}

\begin{document}
	\title{Characterization and space embedding  of directed graphs and social networks through magnetic Laplacians}
	\begin{abstract}
Though commonly found in the real world, directed networks have received relatively less attention from the literature in which concerns their topological and dynamical characteristics.  In this work, we develop a magnetic Laplacian-based framework that can be used for studying directed complex networks.  More specifically, we introduce a specific heat measurement that can help characterizing the network topology.  It is shown that, by using this approach, it is possible to identify the types of several networks, as well as to infer parameters underlying specific network configurations.  Then, we consider the dynamics associated to the magnetic Laplacian as a means of embedding networks into a metric space, allowing the identification of mesoscopic structures in artificial networks or unravel the polarization on political blogosphere.  By defining a coarse-graining procedure in this metric space, we show how to connect the specific heat measurement and the positions of nodes in this space.
	\end{abstract}

\author{Bruno Messias}
\email{messias@ifsc.usp.br}
\affiliation{Instituto de Física de São Carlos, Universidade de São Paulo,  São Carlos, SP 13566-590, Brazil}

\author{Luciano da F. Costa}
\affiliation{Instituto de Física de São Carlos, Universidade de São Paulo,  São Carlos, SP 13566-590, Brazil}
\maketitle

\section{Introduction}

Data structures named graphs (a.k.a. network) can represent several real-world scenarios. Therefore, the analysis of these structures provides useful insights and a deep understanding of phenomena occurring in society or nature, such as in the analysis of crimes involving corruption~\cite{Ribeiro2018} or in the field  of quantum phase transitions~\cite{prl2017quantum2net,quantum2net2018}. Although many of these scenarios can only be faithfully modeled through directed networks~\cite{prx2018},  more effort has been spent in the development of methodologies for undirected networks~\cite{ NaturePhysicsRenormalizaton2018, preModelFitEntropy2018}. Consequently, defining new measures and methodologies devoted to the directed case is of paramount importance for advancing of network science.

Here, using the formalism of magnetic Laplacians~\cite{lieb1993fluxes,preMagCommunity2017}, we define new measures and methodologies which do not ignore the information about direction. In addition, these methodologies have a  conceptual interpretation closely related to the concepts of quantum mechanics. We validate  of these methodologies and measures by showing that they answer the following questions affirmatively:

\begin{enumerate}
\item \label{question1} Can we use the proposed approach to distinguish between networks generated by different models?
\item \label{question2} Is it possible to define an approach capable of inferring parameters of a model that created a certain network?
\item \label{question3} Can we extract information about structure and dynamics, as well the relationships between them?  
\end{enumerate}

If the answer to question \ref{question1} is positive, then measurements related to magnetic Laplacian can be used to classify networks generated by different models. For that purpose, we define a specific heat, $c_\lambda$ , associated to a given graph. We obtain $c_\lambda$  values for several categories of complex networks and also use $c_\lambda$  for training a \emph{Self-Organizing Map} (SOM)~\cite{kohonen1998self,ritter1989self}, which is a type of neural network.  SOM is able to learn which networks belong to the same network type without any previous knowledge. Therefore,  $c_\lambda$  can be understood as a spectral ``fingerprint'' of complex networks.    The relationship between $c_\lambda$  and the network  model resembles those that are found in physical materials, where different types of materials have distinct $c_\lambda$  behaviors.

For undirected networks,  question  \ref{question2} has been addressed previously using an informational point of view~\cite{bianconi2009}, focusing on the entropic distance and Kullback-Leibler divergence~\cite{preModelFitEntropy2018,prx2016} between two networks.  This point of view is extended in the present work for the directed case. The entropic distance or Kullback-Leiber divergence  defined in~\cite{preModelFitEntropy2018,prx2016}  assumes that nodes have identity, \textit{i.e.} permutations of rows and columns of the adjacaceny matrix matters.  However, in case we are interested only on the topology of the networks, the aforementioned methodology will  identify two topologically identical networks, labelled differently, as being distinct.   In addition, it is verified that the entropic distance methods cannot guarantee high accuracy in the estimation of network topological properties for Barabási-Albert networks.  So as to avoid these limitations, we develop a method to measure the dissimilarity between networks based on $c_\lambda$  measures.   As we show here,  this new method achieves low relative errors  (between $0\% $  and $1\%$).

Question \ref{question3} was partially answered by Fanuel \textit{et al.}~\cite{ preMagCommunity2017, magneticT2space2018} and others~\cite{band2014nodal, lieb1993fluxes,Berkolaiko2013}.   In order to improve the classical methods of community detection in directed networks, in~\citep{preMagCommunity2017} the authors defined a correlation measure between distinct vertices by using the formalism of magnetic Laplacian. Additionally, the same research group reported how each vertex of a directed graph could be mapped into a point of a $\mathbb{T}^2$~\cite{magneticT2space2018} space.  By employing this map, the authors were capable of unraveling mesoscale structures (topological) that occurs in real directed networks. The results obtained by Fanuel \textit{et al.}~\cite{ magneticT2space2018} indicated that the magnetic Laplacian comprises information about the structure of directed graphs, providing an affirmative (partial) answer to question  \ref{question3}.  The missing aspect is about the relationships between dynamics and structure. In this work, we address this problem by defining a dynamical system using the quantum evolution operator.  This dynamical system can be used to map a directed (or undirected) graph into a hidden metric space defined by the diffusion.  We show that this new hidden metric space can be used to detect communities in networks, and also to characterize mesoscopic structures in real-world networks, such as those related to political opinion. 

As a mean to connect all concepts developed in this work, we propose a method resembling the geometric renormalization techniques~\citep{kadanoff1966scaling} capable of deriving progressively simplified representations of complex networks. 

This work starts by presenting the basic concepts -- including the definition of the magnetic Laplacian, the statistical mechanics approach employed to graphs, and SOM networks, and follows by reporting developments related to entropy and specific heat inference methods for directed graphs.  Then, we present an embedding procedure capable of mapping vertices of graphs into a hidden space underlain by magnetic Laplacian quantum dynamics.  We then proceed by presenting and discussing the obtained result.  First, we show that the SOM approach is able to identifying the types of complex networks. Subsequently, we show that the inference method, employing $c_\lambda$, exhibits better accuracy than the one via the entropic distance.  The central issue of relating structure and dynamics of complex systems is addressed next, and we show that the geometry defined by quantum dynamics in modular networks has a similar behavior than a frustration-based approach reported previously~\cite{magneticT2space2018}, which is also capable of analyzing undirected graphs.  An illustration of the potential of the quantum dynamics approach is also provided respectively to the classification of nodes in a political opinion network. Finally, we connect the behavior of the specific heat with the embedding of the topology of networks (consequently, with the dynamics associated with magnetic Laplacian) using a coarse-graining procedure.

\section{Methods}

\subsection{Magnetic Laplacian and specific heat measure}

A directed graph can be expressed by a tuple $G=(V, E, w)$, where $V$ is the set of vertices; $E$ is the set of edges such that for each $u, v \in V$  the ordered tuple $e = (u, v) \in E$ assigns a directed edge from vertex $u$ to $v$ and  $w: E\to \mathbb{R}$. A directed graph can be associated with an undirected counterpart  $G^{(s)}=(V , E^{(s)}, w^{(s)})$,  where $w^{(s)}(u, v) =\frac{w(u, v) + w(v, u)}{2}$.   However,  the directionality of  $G$ is lost in $G^{(s)}$. 

In order to preserve the Hermiticity and the information about directionality , define   $\gamma$, as 
$
\gamma: E \to  \mathcal{G},
$
where $\mathcal{G}$  is a group, such that  $\gamma(u, v )^{-1} = \gamma(v, u)$, choosing  $\mathcal{G} = U(1)$  and expressing $\gamma$  as
\begin{align}
\gamma_q(u, v) = \exp( 2  \pi i  q f(u,v)),
\label{eqGamma}
\end{align}
where $q \in [0, 1]$ and   $f(u, v) = w(u, v) - w(v, u)$  represents the  flow in a given vertex $u$ due to another vertex $v$.  

The symmetric graph equiped with $\gamma_q$ has information about directed edges and at the same time the adjacency matrix is Hermitian.

An operator named magnetic Laplacian~\cite{colin2013magnetic,Berkolaiko2013}, $L_q$, can be associated with $(G^{(s)}, \gamma_q)$  . The reason for the term magnetic is that the operator can be used to describe the phenomenology of a quantum particle subject to the action of a magnetic field~\cite{lieb1993fluxes}. Due to this physical context, the  parameter $q$ is named charge.

The magnetic Laplacian, expressed in terms of matrices, is given by
\begin{align}
\mathbf L_q =\mathbf  D -\boldsymbol  \Gamma_q \odot\mathbf  W^{(s)},
\end{align}
where $\mathbf  D $  is the  degree matrix; $[\boldsymbol  \Gamma_q]_{u,v}=[\boldsymbol  \Gamma_q^{\dagger}]_{v,u}=\gamma_q(u, v) $   and $[W^{(s)}]_{u,v} =[W^{(s)}]_{v,u} = w^{(s)}(u, v)$  .

By construction, $\mathbf  D $ and $\mathbf  W^{(s)}$ are both symmetric and  $\boldsymbol  \Gamma_q$  is Hermitian. Consequently,  $\mathbf L_q $ is  Hermitian.  In addition, it is sometimes convenient to use a normalized version of $\mathbf L_q $, which is  given by

\begin{align}
\mathbf  H_q  
= 
\sqrt{\mathbf D^{-1}}\mathbf L_q \sqrt{\mathbf D^{-1}},
\label{eqNormedH}
\end{align}
where the $\mathbf  H_q$ is defined only if   the graph is at least  weakly connected.


A given eigenvector of $\mathbf H_q$, $|\psi_l(q)\rangle\ $,  can be obtained as solution of
\begin{align}
\mathbf H_q |\psi_{l,q}\rangle = \lambda_{l,q}|\psi_{l,q}\rangle
\label{eqEigenVec}
\end{align}

where  $\lambda_{l,q}\in \mathbb R$ and $\lambda_{1,q}\le\lambda_{l,q}\leq\dots\leq\lambda_{|V|,q}$

It is possible to enhance the analogy with physical systems by including a temperature parameter $T \in \mathbb{R}_{>0}$.  By using this parameter, the graph properties can be studied from the statistical mechanics viewpoint.

Here, we  adopted the Boltzmann-Gibbs statistical mechanics formulation as a mens to associate the  partition function

\begin{align}
Z(T, q)= \sum\limits_{l=1}^{|V|}e^{-\frac{\lambda_{l,q}}{T}}
\label{eqPartition}
\end{align}
with $G$.

By using~\eqreff{eqPartition}, the expected value at temperature T  of a  operator  $O$ can be expressed in terms of its eigenvalues $\{o_l\}$  as

\begin{align}
\langle O \rangle = \frac{1}{Z(T, q)}\sum\limits_{l=1}^{|V|} e^{-\frac{\lambda_{l,q}}{T}}o_l.
\label{eqExpectedValue}
\end{align}
In this work, we  use  the~\eqreff{eqExpectedValue} to define the measure of specific heat,  $c_\lambda$ ,  associated with a graph.  This novel measure is given by
\begin{align}
c_\lambda(q, T) = \frac{\langle  H_q^2 \rangle- \langle  H_q \rangle^2}{T^2}.
\label{eqDefSpecifcHeat}
\end{align}

Observe that~\eqreff{eqDefSpecifcHeat} has two free parameters, namely $q$ and  $T$.  The $q$ value can be set so as to increase the quality of separability of modular structures of networks (see the discussion in supplemental material \ref{secSupMatChargeAndSeparability}). In addition, the  parameter $T$ is used in order to investigate the multiscale structure of communities in directed networks~\cite{preMagCommunity2017}. Because of this free choice of parameters and, owing to the fact that we have a rotation associated with directed edges ($\gamma_q$), we 
 plot  $c_\lambda$  in two dimensions , setting $2 \pi q$ as the polar coordinate, and $T$ as the radial one.

\subsection{Self-Organized Maps (SOM)}

SOM is a non-supervised approach to pattern recognition which was originally motivated by the way in which visual information is processed in the retina and visual cortex
~\cite{yin2008self}.  It uses a neuronal network associated topographically to the  input space.  During training, each input
 is mapped into the SOM by using some correlation measurement, such as the internal product.  The neuron having the highest activation value is defined as the winner and its weights are modified so as to become more similar with the input data.  The neighboring neurons  are also modified in a similar manner by using a smoothing function .

In this way, the SOM approach involves competition and correlated learning, so that each neuron competes with the others.

The motivation for considering SOM in this work, instead of other neural network approaches derives from the fact that this specific neural network incorporates topographical mapping of the input space~\cite{ritter1989self}.  One interesting feature of SOM is that it is possible to infer information about how the trained database influenced the learning (e.g. in terms of distances between patterns).  This is normally not possible in the Convolution Neural Network approach.

\subsection{Directed Network Model Inference}

Recent works reported how to use entropic measurements to quantify the distance between two undirected graphs~\citep{prx2016,preModelFitEntropy2018}.  The entropy of a graph is derived from the usual Laplacian spectrum (all eigenvalues are real).  By contrast, these measurements cannot be used in the case of directed graphs because the adjacency matrix is not Hermitian. However,  the magnetic Laplacian methodology yields a Hermitian operator
$H_q$,  which is here used to define an entropic measurement for directed graphs.

Recall that a quantum system at finite temperature, $T$, is defined by its respective density matrix, $\rho(T)$~\cite{blum2012density}. For a graph $G$ and charge $q$, this operator can be expressed in terms of the eigenvalues and eigenvectors associated to  $H_q$ as

\begin{align}
\boldsymbol \rho_q =
\frac{1}{Z(T,q)}\sum\limits_{l=1}^{|V|} e^{-\frac{\lambda_{q, l}}{T}}
|\psi_{l,q}\rangle \langle\psi_{l,q}|.
\label{eqDensity}
\end{align}

The previously defined density matrix can be used in order to define measurements associated with a directed (or undirected) graph. In the next subsections, we show how to perform some that new measurements.

\subsubsection{Entropic distance methodology}

There are several ways to define an entropy measure on a graph~\cite{Ginestra2009, Dehmer2011,pre2015}.
Here, we use the magnetic Laplacian to define a von Neumann entropy associated with a given graph $G$ for a given $q$ and $T$. Such entropy is given by

\begin{align}
S(G, q, T) = \mathrm{Tr}\left[
	  \boldsymbol{ \rho}_q(T)\mathrm{Log}\boldsymbol{\rho}_q(T)\right],
      \label{eqEnt}
\end{align}
where $\mathrm{Log}$ is the matrix logarithm and $\mathrm{Tr}$ corresponds to the trace operation.

Given two  graphs  $\tilde G$  and $G$  using~\eqreff{eqEnt}, here we define an extension for the entropic distance~\cite{prx2016, preModelFitEntropy2018} between the two directed graphs as

\begin{align}
S_d(\tilde G, G, q, T) = 
S(\tilde G, q, T)-\mathrm{Tr}\left[\boldsymbol{\tilde { \rho}}_q(T) \mathrm{Log} \boldsymbol{\rho}_q(T)
\right].
\label{eqDistEnt}
\end{align}

The task of inferring the parameter of an observed graph $\tilde G$, can be posed as finding the parameter $p_{min} \in \{p_i\}$
that minimizes the entropic distance given a number $N_{exp}$ of realizations for each value of $p$,

\begin{align}
p_{min}(q, T) = \textrm{min}_{p_i} 
\left\{ 
\frac{1}{N_{exp}}
\sum\limits_{j=1}^{N_{exp}}
\sum\limits_{i=1}^{|\{p\}|}
S_d(G, G_j(p_i), q, T)
\right\}.
\label{eqDistEntMin}
\end{align}

Although the entropic distance can be used to infer parameters of a generative model (directed or undirected),  the inferred parameters can be somewhat inaccurate. This is due to the fact that the second term on the right-hand  side of the~\eqreff{eqDistEnt} is not independent of permutations of the adjacency matrix (for a given graph with $|V|$ vertices, there are $|V|!$ possibilities of adjacency matrices) associated with $ \tilde G $. Furthermore, because the majority of the generative algorithms ignores the node indices, two identical graphs created with the same set of parameters,  can  result in different adjacency matrices (see the section \ref{secSupMatInference} of the supplementary material for more details).  Therefore, aiming at letting the entropic distance independent of permutations among the labels of the nodes  and to improve the accuracy, we replace the second right-hand side of~\eqreff{eqDistEnt} by the following term 
\begin{align}
\sum\limits_{m=1}^{|V|}\frac{e^{-\frac{\tilde\lambda_{q, m}}{T}}}{\tilde Z(T,q)}\log \frac{e^{-\frac{\lambda_{q, m}}{T}}}{ Z(T,q)}.
\label{eqTrick}
\end{align}
\subsubsection{Specific heat deviation as a measure of dissimilarity }

The methodology of entropic distance reported in~\cite{prx2016,preModelFitEntropy2018}    has two limitations concerning the inference  of parameters. The first limitation  is that this method is not able to compare networks with different number of nodes. This limitation can be a problem when this method is employed to compare  a weakly-connected graph  and a network generated by an algorithm.  The second limitation consists in the fact that the entropic distance is dependent on permutations of the adjacency matrices associated with the graphs. For instance, two isomorphic graphs  can have distinct adjacency matrices, which results in a non-null entropic distance.  This limitation can be  avoided  by using the substitution presented in~\eqreff{eqTrick}. However,  this substitution causes lack of meaning of the expression  ``entropic''.

Here, we propose a novel method that allows us to compare  two networks avoiding the  previous aforementioned limitations.  This comparison is quantified by  the relative deviation of the respective specific heats associated with the two networks. This relative deviation is given by 

\begin{align}
D(\tilde c_\lambda, c_\lambda, q) = 
\frac{\int\int |\tilde c_\lambda(q, T) - c_\lambda(q, T) |\mathrm d T\mathrm d q}{\int\int \tilde c_\lambda(q, T) \mathrm d T\mathrm d q},
\label{eqDistCv}
\end{align}
where the integration region for charge and temperature can be chosen in arbitrary.

Using~\eqreff{eqDistCv}, the solution for the  inference problem of a parameter $p$  is given by the following  equation

\begin{align}
p_{min}(q) = \textrm{min}_{p_i} 
\left\{ 
\frac{1}{N_{exp}}
\sum\limits_{j=1}^{N_{exp}}
\sum\limits_{i=1}^{|\{p\}|}
D(\tilde c_\lambda, c_\lambda, q) 
\right\}.
\label{eqDistEntMin}
\end{align}

\subsection{Graph Embedding}
Graphs represent the relationships (edges) between entities (vertices).  Such relations can be associated with scalar values (weights), which are not necessarily related to a distance measurement in a metric space.  However,  efforts have been made to derive methodologies aiming to embed vertices of a  graph into some hidden metric space~\cite{krioukov2010hyperbolic, nature2017,arxivDiGraph2018}.  
By doing so, it is possible to infer new insights about the studied graphs, such as vertex similarity, community structure, among other possibilities.  In addition, these embeddings can be employed for several practical applications, resulting in more efficient algorithms, such as node classification, community detection, link prediction~\cite{backstrom2011supervised} and routing algorithms~\cite{NaturePhysicsRenormalizaton2018}.  

Here, we propose an embedding methodology for directed networks through dynamics associated with magnetic Laplacian. Additionally, before presenting this novel methodology, we review the frustration based approach, which was proposed as a mechanism to visualize directed networks~\cite{magneticT2space2018}.

\subsubsection{Magnetic Eigenmaps (Frustration-based)  approach  }
\label{secMethodsFrustation}

In~\cite{magneticT2space2018}, the authors reported how to employ magnetic Eigenmaps to map vertices of a directed graph into points of a
 $\mathbb{S}^1$ space through the solution of angular synchronization problem~\cite{cucuringu2016sync}. This problem consists of determining the set of angles
$\{\phi\}$ that minimize the frustration measurement, which is defined as
    
    \begin{align}
    \eta(\{\phi\}, q) = \frac{1}{2}
        \frac{\sum\limits_{u,v\in E}w^{(s)}(u, v)|1-e^{i(2\pi q f(u, v)- (\phi_u -\phi_v))}|}{\mathrm{vol}(G^{(s)})}
        \label{eqFrustation}
    \end{align}

The $|V|$-phases ($\{\phi_0(1), \phi_0(2), \dots, \phi_0(|V|)\}$) of the first eigenvector ($|\psi_0(q)\rangle$) of the magnetic Laplacian~\eqreff{eqNormedH} approximate the solution of the synchronization problem~\cite{singer2011angular}.  Therefore, each vertex can be associated with a point in a $\mathbb{S}^1$ space.  The phases of the second eigenvector can be used to map the set of the vertices into a set of points in a different $\mathbb{S}^1$ space.  Thus, a directed graph can be embedded into $\mathbb{T}^2=\mathbb{S}^1\times\mathbb{S}^1$~\cite{magneticT2space2018}.  However, this embedding procedure brings no information in the limit of $q=0$  or undirected networks, because for both cases, all vertices are mapped into the same point in the space.

\subsubsection{Dynamics-based approach}

To formulate a method capable of mapping directed or undirected networks into a hidden metric space, we propose a space embedding method based on the dynamics associated with the magnetic Laplacian. 
In order to do so, we build up a dynamical system with respective state space formed by vectors $|\phi\rangle \in \mathbb{C}^{|V|}$, a evolution parameter $s\in \mathbb{R}_{>0}$, and use as evolution operator the  evolution operator employed in quantum mechanics, which is smoothed by the Laplace transform (please refer to supplementary material  \ref{secSupMatEvoOperator}).  This evolution operator is given by

\begin{eqnarray}
\mathcal{U}_q(s) = i\mathbf{G}_q(is) = \sum\limits_{l=1}^{|V|}\frac{|\psi_{l,q}\rangle \langle \psi_{l,q}|}{s+i\lambda_{l,q}}   |
    \label{eqLaplaceEvo}
    \end{eqnarray}
   where $\mathbf{G}_q$ is known as a \emph{propagator} in the literature of many body physics and quantum transport~\cite{bruus2004many, datta1997electronic, 2018physRevEquantumTransportCommu}. When this evolution operator is applied in a localized state in a vertex $u$, which is given by
\begin{align}
[|u\rangle]_i=\langle i | u \rangle =\begin{cases}
0,\  \ \mathrm{if} \   \ i\neq u \\
1,\  \  \mathrm{otherwise} 
\end{cases},
\end{align}
it returns an evolved state 
\begin{align}
|\tilde u_q(s)\rangle = \mathcal{U}_q(s)|u\rangle.
\label{eqULaplace}.
\end{align}

This evolved state captures information about the network structure, and this information can be used to build up the embedding procedure.  In order to filter which part of this information is used, we consider two aspects of magnetic Laplacians and its associated dynamics:  the phase-shift, caused by directed edges, and its relationships with the Arahanov-Bohm effect~\cite{aharonov1959significance}. Both aspects draw our attention to the importance of phases in the physics of directed complex networks. Remarkably, we can use the phases ($\theta: V\to\mathbb S^{|V|}$) associated with the evolved state, ~\eqreff{eqULaplace}, to embed a graph in a hidden space. Due to this embedding, a graph can be associated with a set of points  $\Theta= \{\theta(u)\}$.  Unfortunately, due to the high dimensional characteristic, this embedding is not useful for visualization or many other practical applications.

Regardless of the difficulty of this high-dimension characteristic of phase-space,  we can use the set $\Theta$ associated with that space to derive a new space with fewer dimensions. To accomplish this dimensionality reduction, we use the  Diffusion Maps algorithm~\cite{coifman2006diffusion,faceDiffMaps2013,farbman2010diffusion}. The first step to apply the diffusion map to the phase-space is to define a distance function  between two vertices $u, v\in V$. Here, we use a metric that is represented by

\begin{align}
d_q(u, v)^2 =(\boldsymbol \theta(u)-\boldsymbol\theta(v))^2.
\label{eqDistDiffusion}
\end{align}
Now, using this distance function, we define a similarity measurement between two vertices through a kernel function $k_q:\Theta\times\Theta\to \mathbb R_{\ge 0}$ as

\begin{align}
k_q( \theta(u),  \theta(v)) = 
\exp\left(
-\frac{1}{\epsilon}\frac{d_q(u,v)^2}{(2|V|\pi)^2})
\right),
\label{eqKernelFunc}
\end{align}
where $\epsilon\in \mathbb R_>0$.

As can be noted, the kernel is positivity preserving, $k_q(u,v) \ge 0$. Therefore, we can use that function to define a transition probability from $\mathbf \theta(v)$ to $\mathbf \theta(u)$,
\begin{align}
p(u,v) = \frac{k_q(\theta(u),\theta(v))}{n_{q}(u)},
\end{align}
where $n_{q}(u) = \sum\limits_{l \in V} k_q(\theta(u), \theta(l))$.
This transition probability function can be represented by a transition matrix of a Markov chain on $\Theta$.  This matrix is given by,
\begin{align}
\mathbf P_q = \mathbf N_q^{-1}\mathbf K_q
\end{align}
where  $\mathbf K_q$ is  a $|V|\times|V|$ matrix,  in which   $[\mathbf K_q]_{u,v}=k_q(u, v)$ and $\mathbf N_q$ is   a  $|V|\times|V|$ diagonal matrix, such  $[\mathbf N_q]_{u,u}=n_q(u)$.

Using the transition matrix, the transition probability to go from $\theta(v)$ to $\theta(u)$ through $t$-steps ($t\in \mathbb{N}$)  is
\begin{align}
p(v, t| u) = [\mathbf P_q^t]_{u,v    }.
\end{align}
The previous equation can be used to define a diffusion distance at time $t$  between two vertices as follows
\begin{align}
D_t(u,v)^2 = \sum\limits_{x\in V}(p(x,t|u)-p(x,t|v))^2.
\label{eqDiffDistance0}
\end{align}

The diffusion distance,~\eqreff{eqDiffDistance0}, brings new information about the embedding into $\mathbb S^{|V|}$ space. However, it is still necessary to perform dimensionality reduction. Because the similarity measure is symmetric ($k_q(\theta(u), \theta(v)) = k_q(\theta(v), \theta(u)), \  \  \forall  \  \ u,v \in V$), the diffusion distance can be expressed by the eigenvalues and eigenvectors of  $\mathbf P_q$ as

\begin{align}
D_t(u,v) ^2  = \sum\limits_{i=1}^{|V|} (l_i^t \psi_i(u) -l_i^t \psi_i(v))^2 , 
\label{eqDiffDistance1}
\end{align}
where $l_i$ is the $i$th largest eigenvalue of $\mathbf P_q$ and $\psi_i(u)$ is the $u$th component of $i$th eigenvector of $\mathbf P_q$.
As can be noted from~\eqreff{eqDiffDistance1}, the diffusion distance between two vertices ($u$ and $v$) is equivalent to a Euclidean distance between two points in a $\mathbb R^{|V|}$ space. Therefore, we define the function $R_t:V\to\mathbb R^{|V|}$ as

\begin{align}
\mathbf R_t(u) = \begin{pmatrix}
l_1^t\psi_1(u) \\ 
l_2^t\psi_2(u) \\ 
\vdots \\
l_{|V|}^t\psi_{|V|}(u) \\ 
\end{pmatrix}.
\label{eqDiffDistance2}
\end{align}
In addition, we define a truncated version of $R$ through the function  $ r_t: V \to \mathbb R^{m}$;  where $1 < m \le |V|$. In this way, the matrix representation of $r_t$ for the vertex $u$ is given by
\begin{align}
\mathbf  r_t(u) = \begin{pmatrix}
l_1^t\psi_1(u) \\ 
l_2^t\psi_2(u) \\ 
\vdots \\
l_{m}^t\psi_{m}(u) \\ 
\end{pmatrix}.
\label{eqDiffComponent}
\end{align}

The dimensionality reduction can be employed by considering that  $l_1 \ge l_2 \ge> \dots \ge l_{|V|}$ and adjusting the parameter $\epsilon$ to approximate, with reasonable accuracy, the diffusion distance only with some initial components of $\mathbf r_t$. 

By choosing the appropriates parameters ($\epsilon$, $t$ and $m$) and removing the first coordinate of $r_t$ (the first eigenvector is trivial) through the map $l_i^t \rightarrow 1- l_i^t$, we can embed a graph (directed or undirected) into a lower dimensional Euclidean space.

\section{Results}

\subsection{Question 1 - Classification of directed graphs}
\begin{figure}[!htb]
    \includegraphics[width=\columnwidth,keepaspectratio]{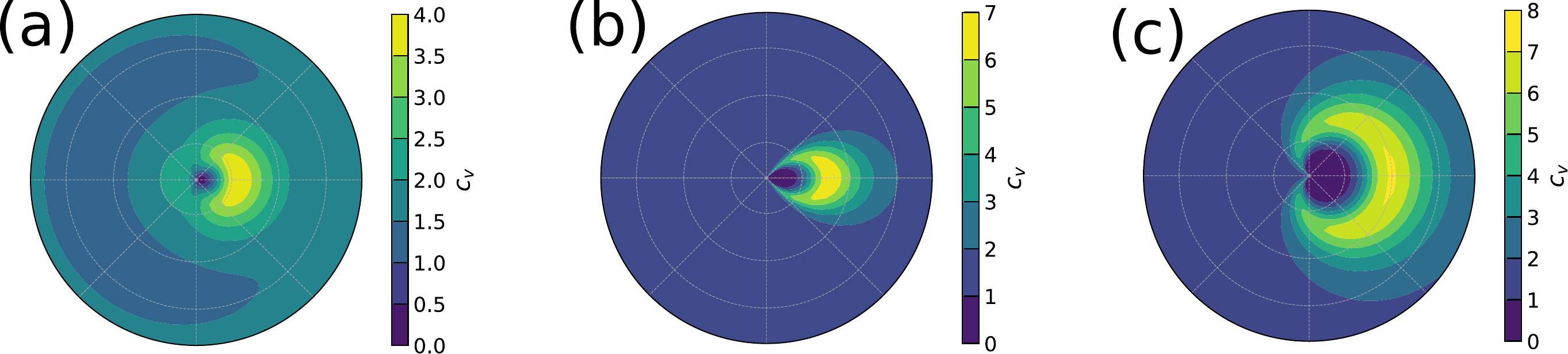}
    \caption{ In (a), (b) and (c) it is shown the specific heat in terms of the charge $2\pi q$ (polar coordinates) and temperature (radial coordinate) for a  scale-free (SF) , Erdős–Rényi (ER) and  Barabási-Albert (BA), network respectivaley. The parameters used for generate those networks was $|V|=1000$; the edge probability for ER was $p=0.003$; the  number of outgoing edges for BA network were $m=3$. The temperature range and charge are uniformly sampled form interval $[0.01, 0.15]$  and $[0, 1/2]$ with $30$ points each. As can be noted the $c_\lambda$ shows a specific pattern for each network. This \emph{fingerprint pattern} for each network explains why the SOM was so successful in the task of organizing networks belonging to the same classes onto the same groups using only the specific heat, without any knowledge about that classes (unsupervised learning).  It follows from~\eqreff{eqGamma} that the eigenvalues, and therefore $c_\lambda$ ,  are symmetric with respect to the addition of integer values to the charge $\gamma_q = \gamma_{q+j}\ \  \forall \ \ j  \in \mathbb{Z}$, reflecting in the bilateral symmetry with respect to the horizontal axis in (a), (b) and (c).}
    \label{figNets2Cv}
\end{figure}

The first question we address refers to the classification of networks (question \ref{question1}): given a  result of a measurement $\mathcal{M}$ on a graph $G$, which generative model created that graph? In this work, we opted to use the specific heat, $c_\lambda$, measurements on graphs in order to answer this question.   As shown in\figref{figNets2Cv}, the  $c_\lambda$  measures yields unique behavior for each generative model. Therefore, $c_\lambda$  provides valuable information which can be used to classify and discriminate between different complex networks types.

\begin{figure}[!htb]
    \centering    \includegraphics[width=\columnwidth,keepaspectratio]{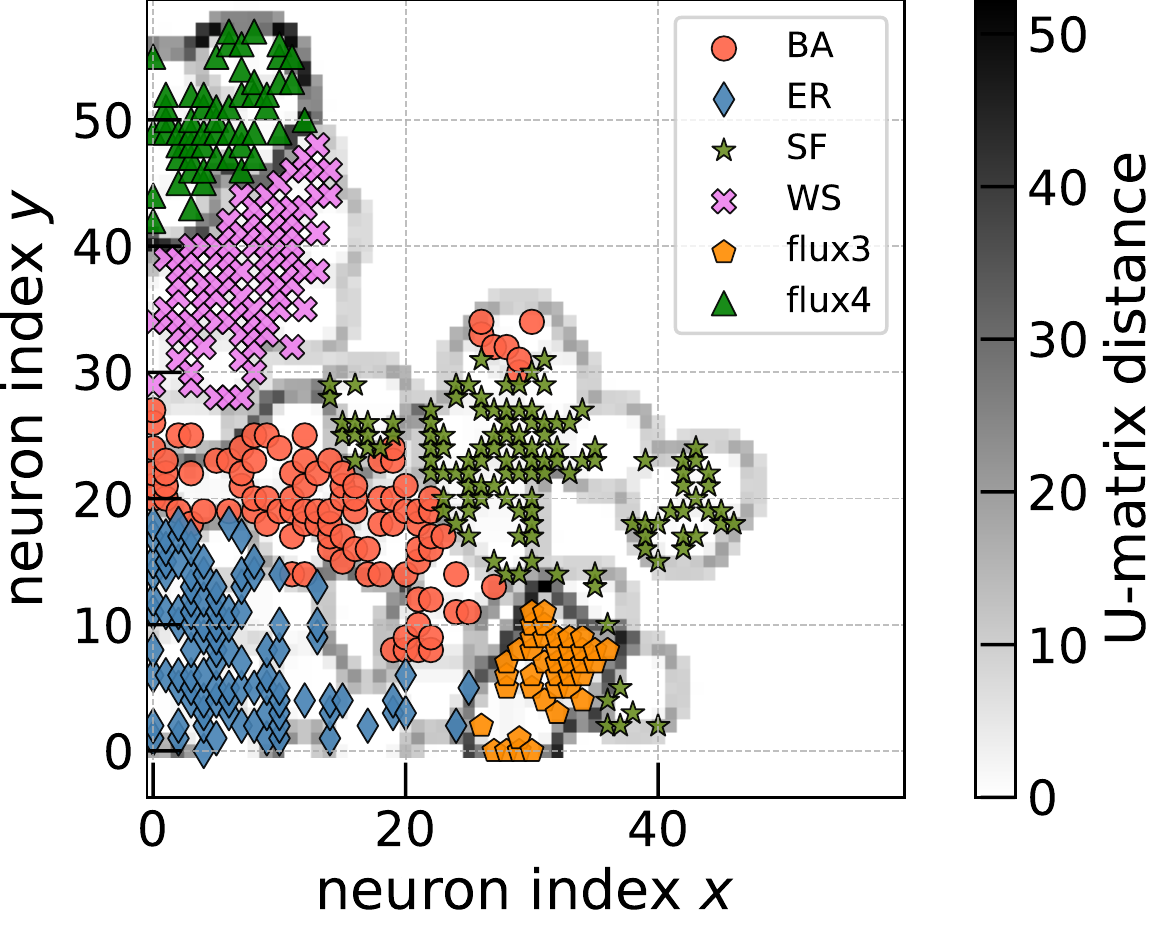}
    \caption{SOM mapping of six types of complex networks represented by the specific heat approach.  The \emph{neuron index x} and \emph{neuron index y}  correspond to indices of neurons in the SOM cortical space.  The distances between neighboring neurons (U-matrix) are indicated in gray.  A good separation between the types of networks can be observed.}
 	\label{figSOMLabels}
\end{figure}

As a mean to evaluate the efficiency of using $c_\lambda$  as a fingerprint of a directed network, we created a dataset with $2000$ $c_\lambda$  samples. This dataset consists of $c_\lambda$  extracted from networks of  the following  types: Erdős–Rényi (ER),  Barabási (BA), scale-free (SF), Watts-Strogatz  (WS),  and modular directed networks with $3$ (flux3) and $4$ (flux4) modules, (please refer to the supplementary material \ref{refSupMatNflux} and \ref{secSupMatDatabase} for more explanation about how the dataset and these modular structures were created). Then, a self-organizing map were trained with these samples, and  the obtained regions  were subsequently labeled. This labeling procedure was done by feeding each training data into the SOM and choosing the neuron that exhibited the highest activation. As indicated by the results,  shown in\figref{figSOMLabels},  networks belonging to the same class have been mapped into nearby neurons, defining respective clusters.  So, the SOM was able to find the patterns of $c_\lambda$  associated to the considered networks without considering previous knowledge (unsupervised recognition). Therefore, this results gives a  positive answer to the question
\ref{question1}.

\subsection{Question 2 - Directed Network Model Inference}

The results presented in\figref{figSOMLabels}  attests that, given a network  $G$, we can determine which model generated it. Furthermore, to complete the characterization procedure of a network, it is necessary to find which parameter was used to generate that network.  In this section, we explore this problem by using two methodologies developed in the current work: the extension for entropic distance for directed networks and the relative deviation of the specific heat.

\begin{figure}[!htb]
    \centering
    \includegraphics[keepaspectratio, scale=0.5]{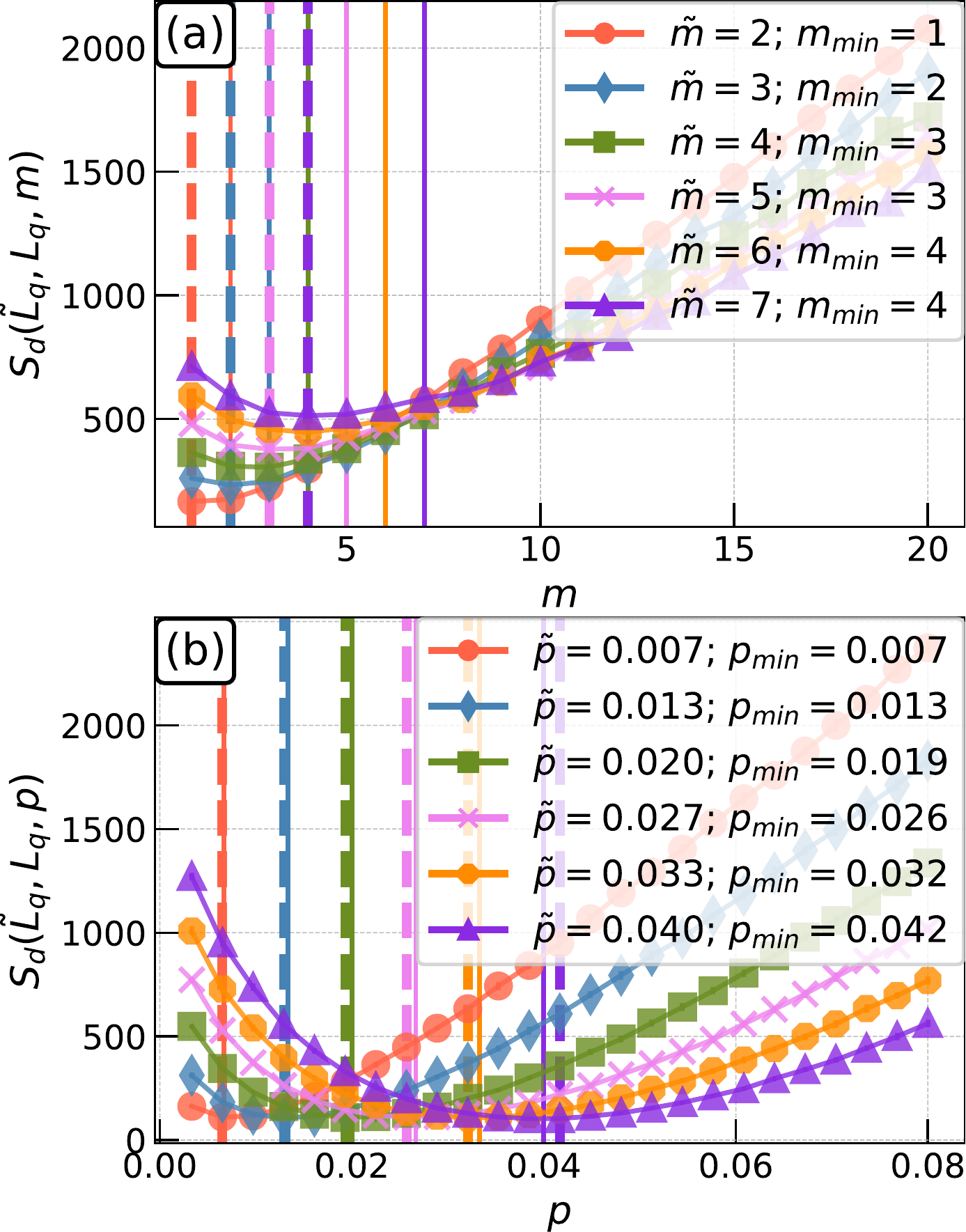}
    \caption{
    The curves shown in (a) and (b) represent the entropic distance in terms of the parameters adopted for network generation, considering $q=1/3$, $N_{exp}=25$, and $|V|=300$ for both ER(b) and BA(a) networks.}
   \label{figDistEnt}
\end{figure}

Figure~\ref{figDistEnt} presents the inference results obtained by applying entropic distance methodology with the trick presented in, ~\eqreff{eqDistEnt}, for  BA (a) and ER (b)  network models.  For each network model, six different networks were created. Each ER and BA networks were generated by using distinct connecting probabilities $\tilde p$ and out-degrees $\tilde m$, respectively. The continuous vertical lines show the correct value of the parameter and the vertical dashed lines identify the position of the minimal entropic distance, which is the inferred value of the parameter. As it is seen in, the estimated results for ER are close to the correct values (around $4\%$ of the relative error). However,  the estimated parameter ($m_{min}$) for BA networks resulted significantly divergent from the original settings (errors larger than $40\%$), indicating that the entropic distance is inadequate for estimating the out-degree parameter of BA model. 

\begin{figure}[!htb]
    \centering
  \includegraphics[scale=0.5,keepaspectratio]{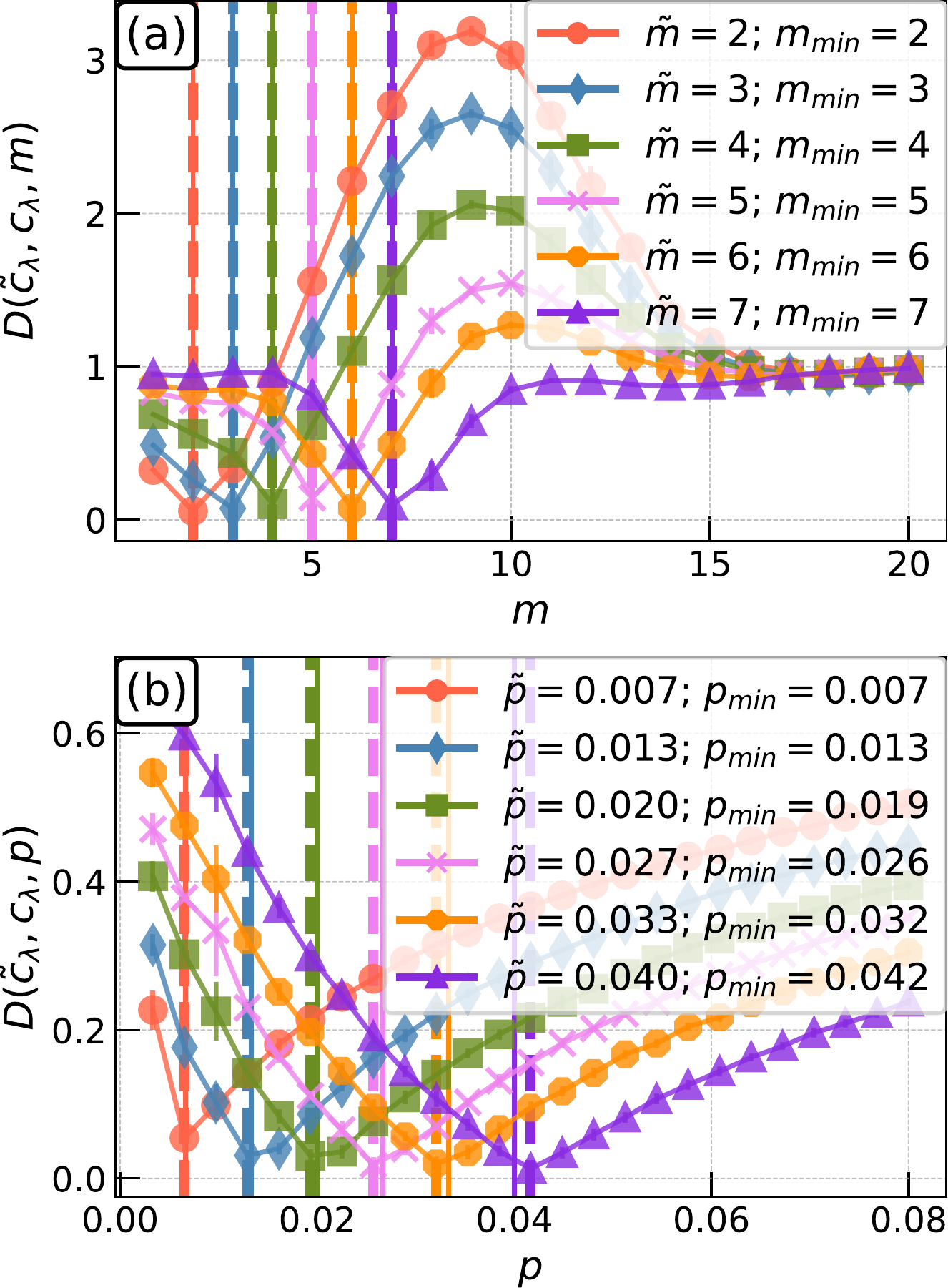}
    \caption{The curves presented in (a) and (b) represent the deviation of  specific heat~\eqreff{eqDistCv}, respectively to BA and ER, in terms of the parameters adopted for network generation, considering $q=1/3$, $N_{exp}=25$, $|V|=300$, and $T$ for 10 points uniformly spaced within the interval [0.01, 0.7].   }
      \label{figDistCv}
\end{figure}

Figure~\ref{figDistCv} presents the inference results when we employ the approach based on the specific heat measurements.  As can be seen, the estimated parameters are close to the respective correct values for ER networks. Remarkably, in contrariwise to the results shown in\figref{figDistEnt} (a), the values inferred ($m_{min}$ ) for the number of outgoing edges for the BA networks, \figref{figDistCv}(b), are exactly matched to the original counterparts.

As can be noted, the global minimum of the deviations of the specific heat measurements is in sharp  contrast to the entropic inference method, which yields a flat minima.  This distinct behavior benefits to infer the parameters because the minimum values tend to be more evident.

\subsection{Question 3 - Structure and Dynamics}

In this section we address the third question motivating the current work, namely the relationship between structure and dynamics in directed complex networks. Here, we propose that this type of relationships can be unraveled by exploiting the geometry brought about by quantum dynamics  and diffusion coordinates system. 

However, before studying the dynamics-structure relationships we aim at connecting the geometry implied by the quantum dynamics  (which is proposed in the present work) with the geometry induced by the frustration method~\cite{magneticT2space2018}.  In order to  do so we created a modular (flux3) structure with $N_f=3$, $N_c=100$,  $p_c=0.5$  and set  $p_d=0.8$. Then, we constructed the  magnetic Laplacian associated with that network by adopting $q=1/3$.  

\begin{figure}[!htb]
    \centering
    \includegraphics[scale=0.5,keepaspectratio]{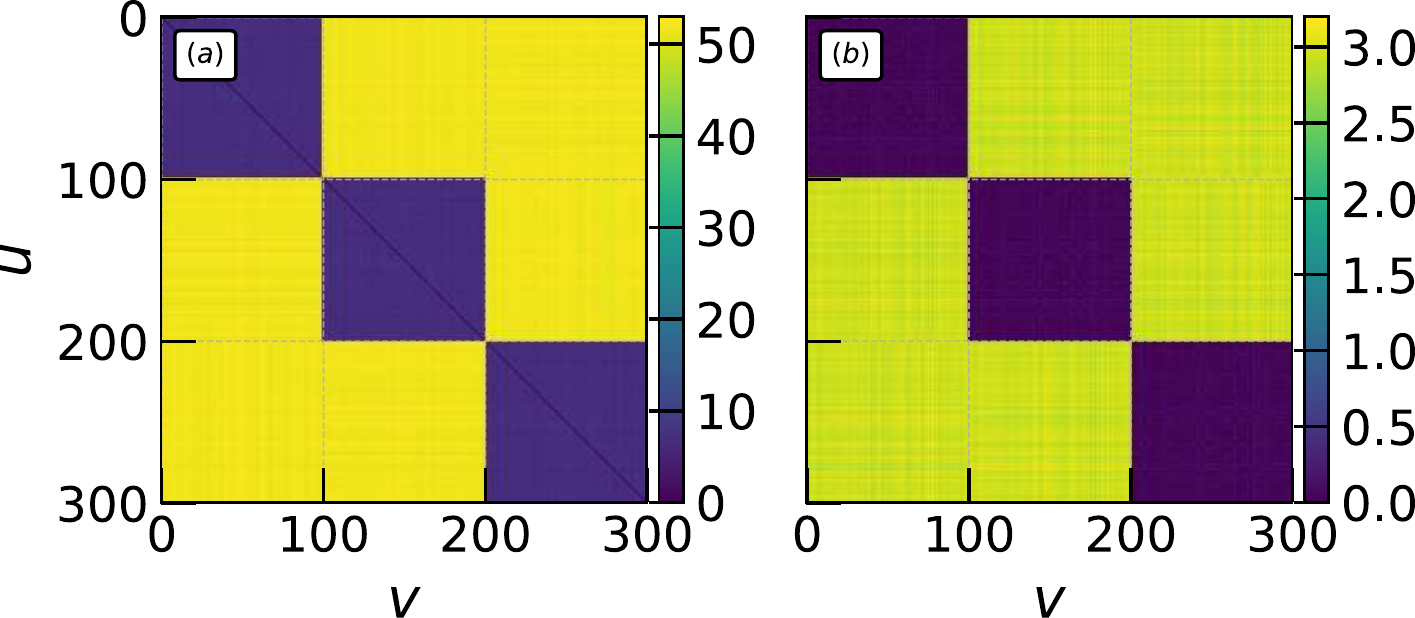}
    \caption{Distance maps between each pair of nodes, $(u, v)$, bellowing to a modular directed network created with $N_f=3$, $N_c=100$ $p_d=0.8$, and $p_c=0.5$.  In addition, we set $q=1/3$ for
all cases.  In $(a)$ it is shown the distance map obtained through equations~\eqreff{eqULaplace} and~\eqreff{eqDistDiffusion} with $s=0.001$ ($\Delta t=10^3$);  $(b)$ gives the distances  evaluated using the phases of the first eigenvector of $\mathbf H_q$. }
    \label{figNf3DynamicsMetric}
\end{figure}

By using~\eqreff{eqULaplace} with $s=0.01$, we  calculated the distance map between all nodes ($u, v$) assuming the metric defined by~\eqreff{eqDistDiffusion}. The results are shown in\figref{figNf3DynamicsMetric}(a)  and represent the distance between nodes embedded in phase space associated with quantum dynamics.  In addition  the results for frustration embedding  are shown in\figref{figNf3DynamicsMetric}(b). As can be noticed, the same pattern of distance was obtained as before\figref{figNf3DynamicsMetric}(a).
The distance maps indicate that if the nodes are separated by short distances in the frustration space ($\mathbb{S}^1$), these nodes will have a similar dynamic behavior  with respect to the phases space ($\mathbb{S}^{|V|}$).  Therefore, for  modular directed networks, the embedding generated by frustration and the embedding generated by dynamics are equivalent. However, we stress that the geometrization implemented by dynamics is based on a concept distinct from that underlying the frustration method. Therefore, this complementary approach has the benefit of providing different information about a given  complex network.

\begin{figure*}
  \includegraphics[scale=0.8]{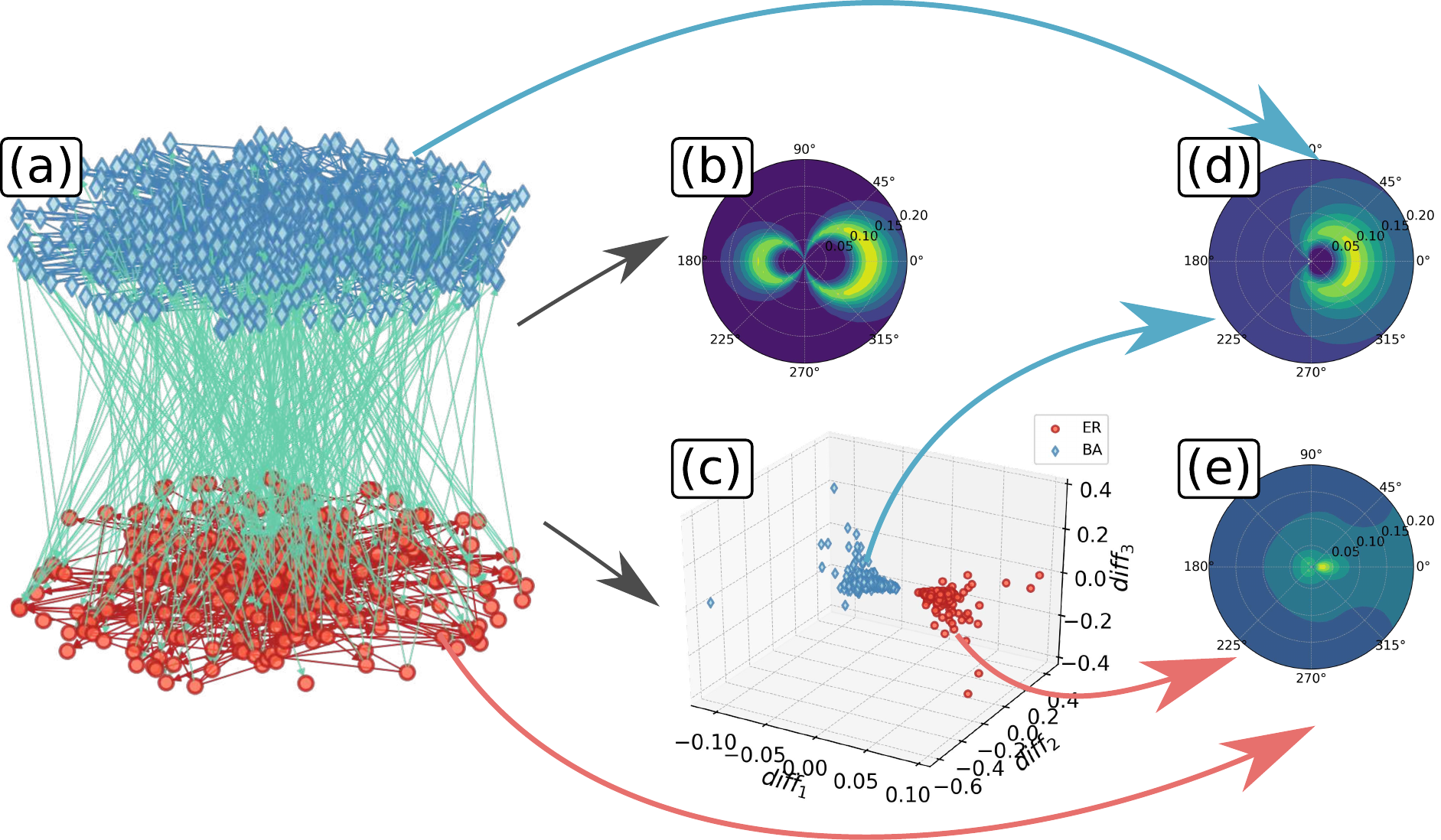}
  \caption{A directed multilayer network ($ G_M $) formed by two simple networks (a), being a network BA (blue diamond) generated with $ m = 3 $ and an ER  network (red circle) generated with $ p = 0.001 $. The vertices of each network were connected randomly with a probability of $ 10 \% $.  (b) shows the $c_\lambda$ associated with $ G_M $, and the embedding of the vertices of $ G_M $ into $ \mathbb {R} ^ 3 $ is shown in (c), where the coordinates are determined by the diffusion map.  The specific heat of each cluster in $\mathbb R^3$  it is shown in (d) and (e).}
  \label{figMultiLayer}
\end{figure*}


The previous results indicates that the space embedding procedure can be employed to detect modular structures in complex networks. Therefore, it is reasonable to use the same framework while studying multilayer networks. Here, we focus on the task of inferring the layers which take part in a multilayer network using the embedding procedure. More specifically, we constructed a two-layered network where one layer was a BA network (Fig. \ref{figMultiLayer}(a) blue diamonds) constructed with $500$ nodes and $m = 3$ and the other layer was  an ER network (Fig. \ref{figMultiLayer}(a) red circles) constructed with $p = 0.005$ and $300$ nodes. After the generation of the networks, the layers were randomly connected by using a probability of $10 \%$ for each pair of nodes belonging to different layers. Observe these edges are directed, meaning that $e(u,v)  \neq e(v,u)$. Subsequently, the specific heat map of the multilayer network was extracted and is presented in \figref{figMultiLayer}(b). In order to realize the embedding procedure, instead of using the similarity matrix given by~\eqreff{eqKernelFunc}, we  opted for using the matrix given by

\begin{align}
\mathbf K_q = \mathbf{H}_q \odot \mathbf{H}_q^{T} .
\label{eqFineMatrix}
\end{align}  
Note that this matrix implies in a much lower computational cost when compared with the matrix presented in~\eqreff{eqKernelFunc}.

We used the matrix given by~\eqreff{eqFineMatrix} and applied the diffusion mapping framework. This framework maps the nodes of the multilayer network into points of ~$\mathbb R^3 $ space, as shown in~\figref{figMultiLayer}(c). In the following, we applied a hierarchical clustering process in these points, which resulted in two clusters. Remarkably, the nodes of the layers were correctly separated into different clusters. Therefore, the specific heat of each cluster, see~\figref{figMultiLayer}(d) and~\figref{figMultiLayer}(e),  represents the specific heat of each particular layer, exhibiting intrinsically different organizations.

\begin{figure}[!htb]
    \centering
  \includegraphics[scale=0.5,keepaspectratio]{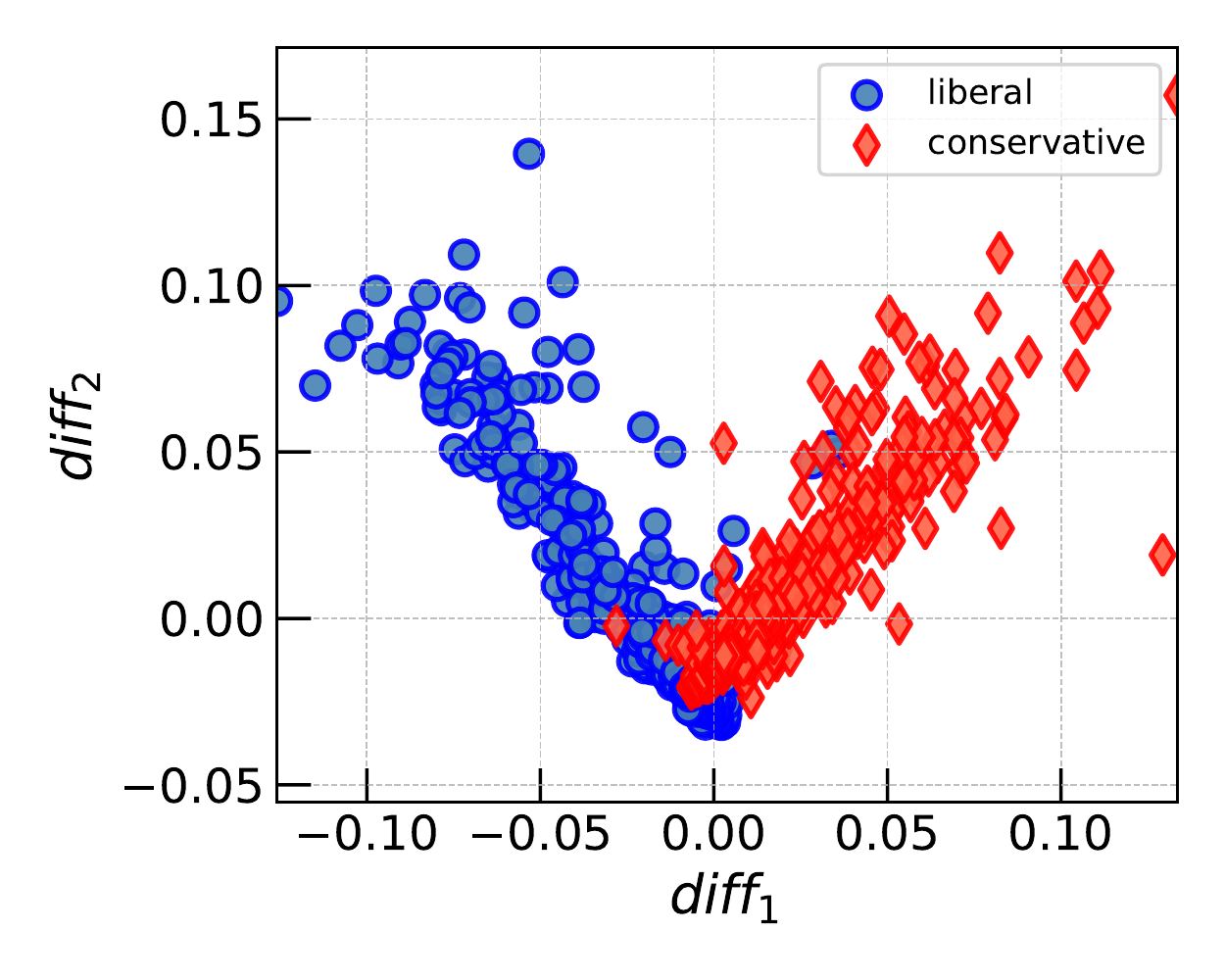}
    \caption{Embedding of the American political blogsphere into the space defined by the dynamics associated to the magnetic Laplacian. The axes correspond to the first (diff1) and second (diff2) components associated to the diffusion map~\eqreff{eqDiffComponent}.}
    \label{figPolitic}
\end{figure}

Having shown that the dynamics of the magnetic Laplacian bears relationships with the abstract frustration approach and can be used to unravel community structures present in artificial networks, we use  the same
methodology  to unravel patterns in real
world networks. As a means to illustrate this possibility, we used a social network derived from the 2005 USA politic blogosphere~\cite{adamic2005political} . 
In this network, the edges correspond to hyperlinks between blogs, while the vertices indicate the political trends of the blogs (blue circles = liberal and red diamonds = conservative).  In\figref{figPolitic}, it is presented the embedding of this network in the diffusion induced space, following~\eqreff{eqDiffComponent}.  The obtained mapping allows a good separation between the two types of opinion.

The results presented in\figref{figNf3DynamicsMetric} and\figref{figPolitic}   
corroborate the hypothesis implied by question \ref{question3}.  More specifically, by using a physical analog (particle dynamics), we can define a geometry associated to a diffusive process so that this structure is capable of revealing structural and dynamical patterns in directed graphs.

\subsection{Connecting concepts  through geometric coarse-graining procedure}

Going back to the three main questions motivating our approach, we observe that questions \ref{question1} and \ref{question2} are not immediately related to question
\ref{question3}.  

However, it is possible to use a coarse-graining procedure in the adopted hidden spaces so as to derive a relationship among spectrum, structure and dynamics in complex networks using hidden metric spaces as a intermediary concept,  as explained in the following. 

\begin{figure}[!htb]
    \centering
    \includegraphics[width=\columnwidth,keepaspectratio]{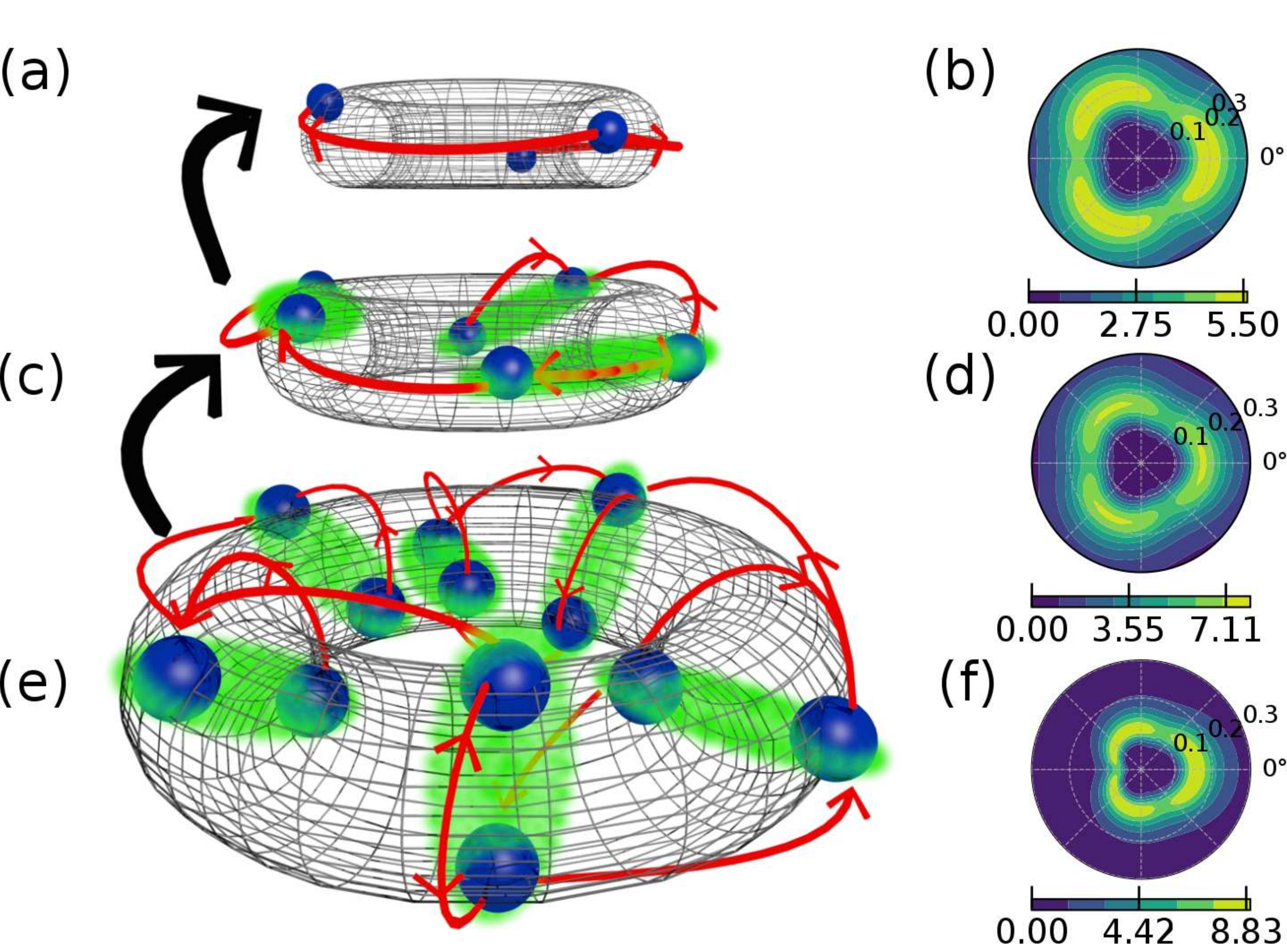}
    \caption{Visual representation (just for didactic purposes) of the coarse-graining procedure is shown in (a), (c) and (e).  Specific heats for each coarse graining-porcedure step are presented in (f), (d) and (b) for a modular directed network with three modules.  Observe that the petals structure is preserved by the renormalization procedure. }
    \label{figCvEVO3FLUX}
\end{figure}

Let us consider the problem of studying how a measurement $\mathcal{M}$ on a graph $G^{(0)}=(V, E, w)$ is related to the adopted hidden spaces, and consequently with the respective dynamics whit generates the embedding into this space.  First, we map $G$ into a metric space $S$, so that each vertex $u\in V$ becomes associated with a point $\mathbf r^{(0)}_u \in S$.  Therefore, the embedding yielding a set of points $\{r^{(0)}_u\}$, as shown in\figref{figCvEVO3FLUX}(e). By construction, it follows that the proximity between vertices in this space is related to the similarity of some  property (e.g.~frustration or dynamics in phase space).
Now, we can define a coarse-graining procedure resembling those used in statistical mechanics~\cite{kadanoff1966scaling,NaturePhysicsRenormalizaton2018}. 

After make the measurement of $\mathcal{M}$ in $G^{(0)}$ , $\mathcal M^{(0)}$, and embedding the graph in the hidden  space,  $\{r^{(0)}_u\}$ we  group vertices  into super-vertices such that the sum of the distances (some metric  must be used, here we are using the Euclidean metric)  between its constituent elements be as small as possible. Each super-vertex represents a vertex of a new graph  $G^{(1)}= (\frac{|V|}{2}, E^{(1)}, w^{(1)} )$, as illustrated in Figure\figref{figCvEVO3FLUX}(c); where $E^{(1)}$ and $w^{(1)}$ 
can be defined so as to preserve some specific property of the original network.  Now, we calculate $\mathcal M$ in $G^{(1)}$, $\mathcal{M}^{(1)}$, and map $G^{(1)}$ into $S$, defining a new set of positions associated to the vertices $\{\mathbf r^{(1)}_u\}$. This renormalization procedure can be repeated for a given number of stages, as shown in\figref{figCvEVO3FLUX}(a).  At the end,  this coarse grained procedure provides the flow
\begin{align}
\begin{rcases}
  G^{(0)} \\
  \mathcal{M}^{(0)} \\
  \{r^{(0)}_u\}
\end{rcases}
\mapsto
\begin{rcases}
  G^{(1)} \\
  \mathcal{M}^{(1)} \\
  \{r^{(1)}_u\}
\end{rcases}
\mapsto
\dots
\mapsto
\begin{rcases}
  G^{(N)} \\
  \mathcal{M}^{(N)} \\
  \{r^{(N)}_u\}
\end{rcases},
\end{align}

As a proof of concept of using aforementioned coarse-graining procedure we choose to better understand the relationship between the specific heat measurement and the hidden spaces of modular directed networks.  In order to do so, we created a  modular directed network  with $N_f=3$, $p_c=50\%$, $p_d=70\%$ and $|V|=300$. The mapping  $c_\lambda$  of this network is shown in\figref{figCvEVO3FLUX}(f). The petal structures obtained for the specific heat are a consequence of modular structures (communities) in the network, as discussed in the Supplementary Material \ref{secSupMatStructureAndCv}. Therefore, we  expect that petal structures  must be preserved during coarse-graining procedure if this structures is correlated with  the positions in the hidden metric space. In other words, if the behavior of $c_\lambda$  is correlated with the macrostrutures in the hidden spaces, we expect then lows scales to vanish in this space, and, if connections among super-vertices are preserved, the behavior of $c_\lambda$ also remains unchanged.

Recall that, as seen in\figref{figNf3DynamicsMetric}, the distances in the frustration-related space and in dynamics-based space present the same pattern.  Consequently, it suffices to study the coarse-graining effect in just one of these spaces.

We applied the coarse-graining method so as to preserve the edges between the super-vertices.  Therefore, the presence of an edge from super-vertex $l$ to $m$ in step $t+1$ is

\begin{align}
w(l, m, t+1) = \begin{cases}
1, &\text{if }  \sum\limits_{\substack{u\in l \\ v\in m}} w(u,v,t)\ge 1\\
0, & \text{orterwise. }
\end{cases}
\end{align}

Using $q=1/3$ (please refer to the supplementary material \ref{secSupMatChargeAndSeparability} for reasons of choose this specific value of charge) we mapped the original graph into the space $\mathbb{T}^2$ and applied the coarse-graining procedure twice.  The specific heat obtained for the effective networks $G^{(1)}$ e $G^{(2)}$ is illustrated in\figref{figCvEVO3FLUX} (d) and\figref{figCvEVO3FLUX} (b), respectively.  As it can be observed, the specific heat behavior (petals structure) was preserved.  Thus, for this type of network, we conclude that the $c_\lambda$   is correlated with the structures formed in the hidden metric spaces.

\section{Conclusions}

In the present work, we reported an approach to study directed networks that is based on concepts and measurements commonly used for materials characterization and transport in condensed matter.  First, we proposed a new measurement related to the spectrum of a graph, namely the specific heat $c_\lambda$.  We postulate that this measurement has a potential to identify network types and to infer respective parameters, which was confirmed in the reported experimental results.  In particular, indications have been provided that this approach to graph characterization can be more accurate than by using entropic distances.  

A graph-embedding approach based on quantum dynamics associated with magnetic Laplacian was also described.  Experimental results have shown that this approach can reveal mesoscopic structure in both theoretical and real-world networks (a political opinion dataset).  We also showed how the \emph{geometric} structures in this hidden spaces can be associated with the specific heat according to a coarse-graining procedure.

Our contributions pave the way to a number of future developments and applications involving directed complex networks.   For instance, these methods can be immediately applied to study several other theoretical and real-world structures, including fake news dissemination, metabolic networks, neuronal systems, to name but a few possibilities.  Regarding the more conceptual aspects of the reported contributions, it would be interesting to consider alternative coarse-graining methodologies so as to preserve specific topology or dynamical behaviors in directed networks. It would also be interesting would be to consider geometrizations defined by dynamics that naturally incorporate temperature as a parameter.

\section*{Acknowledgements}
The authors thank Thomas Peron, Henrique F. de Arruda and Paulo J. P. Souza for all suggestions and  useful discussions. The authors acknowledge financial support from Capes-Brazil, São Paulo Research Foundation (FAPESP)
(grant no.  11/50761-2 and 2015/22308-2), CNPq-Brazil (grant no. 307333/2013-2) and NAP-PRP-USP. Research carried out using the computational resources of the Center for Mathematical Sciences Applied to Industry (CeMEAI) funded by FAPESP (grant 2013/07375-0).

\bibliography{biblio.bib}
\cleardoublepage
	\appendix

    \section{ Directed  modular networks (flux$N_f$)}
\label{refSupMatNflux}
In this work, a directed modular network (flux$N_f$) is a random graph with $|V|$ vertex   created with the following procedures: 
\begin{enumerate}
\item Split the set $V$ onto $N_f$ equal-size sets ($f_1, f_2,\dots,f_{N_f}$), each set representing a community structure. 
\item For each $u, v \in f_i$ create a directed edge $(u, v)$ with probability $p_c$.
\item For each $u \in f_i$ and a $v \in f_{i+1}$ (assuming $f_{N_f+1}=f_1$), create a directed edge $(u, v)$ with probability $p_d$. 
\end{enumerate}

\section{Community structures in network and your relationship with specific heat and symmetries}\label{secSupMatStructureAndCv}
\label{sectionAnaliseDeFluxo}

This section aims at studying the influence of community  structures in modular directed networks on the magnetic Laplacian spectrum and, consequently, on the specific heat, $c_\lambda$. We assume that the connections within the communities, $\mathbf{W_{in}}$, as well as between the communities, $\mathbf{W_{out}}$, are not differentiated between the structures.  Under this hypothesis, the adjacency matrix can be organized as follows, assuming  $N_f$ modules (henceforth, we take $N_f>2$):

\begin{eqnarray}
\mathbf{W} = \begin{bmatrix}
\mathbf{W_{in}}& \mathbf{W_{out}}  & \mathbf{0}_{N_c}& \dots &\mathbf{0}_{N_c}\\ 
\mathbf{0}_{N_c}& \mathbf{W_{in}} &  \mathbf{W_{out}}& \dots&\mathbf{0}_{N_c} \\ 
\vdots & \vdots &  \vdots& \ddots&\vdots \\ 
\mathbf{W_{out}}&  \mathbf{0}_{N_c}&   \mathbf{0}_{N_c}&\dots&\mathbf{W_{in}}
\end{bmatrix},
\end{eqnarray}
where $\mathbf 0_{N_c}$ is a null matrix $N_c\times N_c$.  For generality's sake $\mathbf{W_{in}}$ and $\mathbf{W_{out}}$ can be constructed in arbitrary form.

The magnetic Laplacian expressed as discussed above has the following organization:

\begin{eqnarray}
\mathbf{H}_q = \begin{bmatrix}
\mathbf{H_{in}}& \mathbf{H_{out}}   & \mathbf{0}_{N_c}& \dots &\mathbf{H_{out}}^\dagger\\ 
\mathbf{H_{out}}^\dagger & \mathbf{H_{in}} &   \mathbf{H_{out}}  & \dots&\mathbf{0}_{N_c} \\ 
\vdots & \vdots &  \vdots& \ddots&\vdots \\ 
 \mathbf{H_{out}}  &  \mathbf{0}_{N_c}&   \mathbf{0}_{N_c}&\dots&\mathbf{H_{in}}
\end{bmatrix},
\label{eqHamilOrig}
\end{eqnarray}
Note that this matrix is circulant, i.e.
\begin{eqnarray}
\mathbf{H}_q = \begin{bmatrix}
\mathbf{h}_0& \mathbf{h}_1    & \dots &\mathbf{h}_{N_f - 1}\\ 
\mathbf{h}_{N_f - 1}& \mathbf{h}_{0} &    \dots&\mathbf{h}_{N_f - 2} \\ 
\vdots & \vdots &  \ddots&\vdots \\ 
\mathbf{h}_1   &  \mathbf{h}_{2} &\dots&\mathbf{h}_0
\end{bmatrix}.
\end{eqnarray}
Observe that $\mathbf{H}_q$ is also Hermitian, corresponding to a type of Toepltiz matrix~\cite{gray2006toeplitz}, so that the spectral solutions can be obtained analytically considering the property that all the columns in the original matrix can be expressed as cyclic permutations of the first column.

Our objective now is to find the set $\{\lambda_u\}$ such that  
\begin{eqnarray}
\mathbf{H}_q |\psi_u\rangle = \lambda_u |\psi_u\rangle.
\label{eqAutoVec}
\end{eqnarray}

The eigenvectors can be obtained as
\begin{align}
|\psi_u\rangle= \begin{bmatrix}
 |\phi\rangle\\ 
\rho_u |\phi\rangle \\ 
\vdots \\ 
\rho_u^{N_f -1} |\phi\rangle
\end{bmatrix},
\label{eqEigenVec}
\end{align}
where  $u \in \{0,\dots, N_f-1\}$ and $\rho_u=\rho_{N_f-u}^\star= \exp(\frac{2\pi i u}{N_f})$.
Plugging this eigenvector~\eqreff{eqEigenVec} into~\eqreff{eqAutoVec}, it is enough to solve the block equation induced by the first row
\begin{align}
\mathbf{\tilde H}_u|\psi_u\rangle=\sum\limits_{l=0}^{N_f-1}\mathbf{h}_l\rho_{l\cdot u} |\psi_u\rangle = \lambda_u |\psi_u\rangle,
\end{align}

The above equation can be simplified by taking into account that $\mathbf{H_N}$ is Hermitian $\mathbf h_j = \mathbf h_{N_f - j}^\dagger$.

Introducing the variable
\begin{eqnarray}
m_f =\begin{cases}
\frac{N_f + 1}{2} \text{ if } N_f \text{ is odd},\\
\frac{N_f}{2} \text{ if } N_f \text{ is even}
\end{cases},
\end{eqnarray}
 we can obtain
\begin{align}
\mathbf{\tilde H}_u=\mathbf{h}_0
+\sum\limits_{l=1}^{m_f-1}
    \left(
    \mathbf{h}_l\rho_{l\cdot u} +
    \mathbf{h}_l^\dagger\rho_{l\cdot u}^\star
    \right)+ \mathbf{\Delta},
\end{align}
where
\begin{align}
\mathbf{\Delta}= \begin{cases}
 \mathbf{0}_{N_c}\text{ if } N_f \text{ is odd},\\
 (-1)^u\mathbf{h}_{m_f} \text{ if } N_f \text{ is even}
 \end{cases}.
\end{align}

Since the flow structure $\mathbf{\Delta}=\mathbf{0}_{N_c}$ and only three instances $\mathbf{h}_{u}$ are non-null, we have
\begin{align}
\mathbf{\tilde H}_u=\mathbf{h}_0
+
\mathbf{h}_1\rho_{u} +
\mathbf{h}_1^\dagger\rho_{u}^\star,
\end{align}
Replacing the operators $\mathbf{h}$ by their respective counterparts in equation~\eqreff{eqHamilOrig} we obtain the following expression for the $u$-th matrix, 
\begin{align}
\mathbf{\tilde H}_u=\mathbf{H_{in}}
+
e^{\frac{2\pi i u}{N_f}}\mathbf{H_{out}} +
e^{-\frac{2\pi i u}{N_f}}\mathbf{H_{out}}^\dagger.
\label{eqFluxGeral}
\end{align}
In the following sections we will investigate how different $\mathbf{H_{in}}$ influence $c_\lambda$.

\subsection{Uniform Connections}
Uniform connection is characterized by having the degree of each vertex given as
$[\mathbf D_{ii}] = d = 2N_c - 1$.
Consequently, the intrablock of the magnetic Laplacian is  
\begin{eqnarray}
\mathbf{H_{in}} =\frac{\mathbf{I}_{N_c}(1+d)-\mathbf{1}_{N_c}}{d},
\end{eqnarray}
and the interblock defining the connections between the above structures is given as
\begin{eqnarray}
\mathbf{H_{out}} = -\frac{\exp(2\pi i q)}{2d}\mathbf{1}_{N_c}.
\end{eqnarray}
Plugging the two previous equations into~\eqreff{eqFluxGeral}, 
 $\mathbf{\tilde H}_u$ can be obtained as
\begin{align}
\mathbf{\tilde H}_u&= 
 \frac{\mathbf{I}_{N_c}(1+d)-\mathbf{1}_{N_c}}{d}\nonumber\\
 & -2\frac{\cos(2\pi( \frac{u}{N_f} -q))}{2d}\mathbf{1}_{N_c},
\end{align}
observe that $\mathbf{\tilde H}_u$ is a circulant matrix.   Let $v\in\{0,...,N_c-1\}$, and defining 
\begin{eqnarray}
m_c =\begin{cases}
\frac{N_c + 1}{2} \text{ if } N_c \text{ is odd},\\
\frac{N_c}{2} \text{ if } N_c \text{ is even}
\end{cases},
\end{eqnarray}
the eigenvalues of  $\mathbf{\tilde H}_u$ are obtained as

\begin{align}
\lambda_{u,v}=h_0
+\sum\limits_{l=1}^{m_c-1}
\left(
h_l\rho_{l\cdot v} +
h_l^\dagger\rho_{l\cdot v}^\star
\right)+ \Delta.
\label{eqLambdaUniforme}
\end{align}
where
\begin{align}
\Delta= \begin{cases}
0\text{ if } N_c \text{ is odd},\\
(-1)^v h_{m_c} \text{ if } N_c \text{ is even}
\end{cases}.
\end{align}
Replacing $h_l$ by their counterparts in~\eqreff{eqLambdaUniforme} the following eigenvalue
equation can be obtained
\begin{align}
\lambda_{u,v}&=1 -\frac{\cos(2\pi( \frac{u}{N_f} -q))}{d}\nonumber\\
&+\frac{2}{d}\left(
1+\cos(2\pi( \frac{u}{N_f} -q))
\right)f(v, N_c, m_c)
+ \Delta,
\label{eqLambdaFinal}
\end{align}
where $f(v, N_c, m_c) = \sum\limits_{l=1}^{m_c-1}
\cos(\frac{2\pi v l}{N_c})$, such that
\begin{align}
f(v, N_c, m_c)= \begin{cases}
m_c\text{ if } v=0,\\
\frac{\sin(\frac{\pi v m_c}{N_c}) }{\sin(\frac{\pi v }{N_c})}\cos(\frac{\pi v }{N_c}(m_c-1)) \text{ otherwise}
\end{cases}.
\end{align}

\begin{figure}[!htb]
    \centering
    \includegraphics[scale=.3]{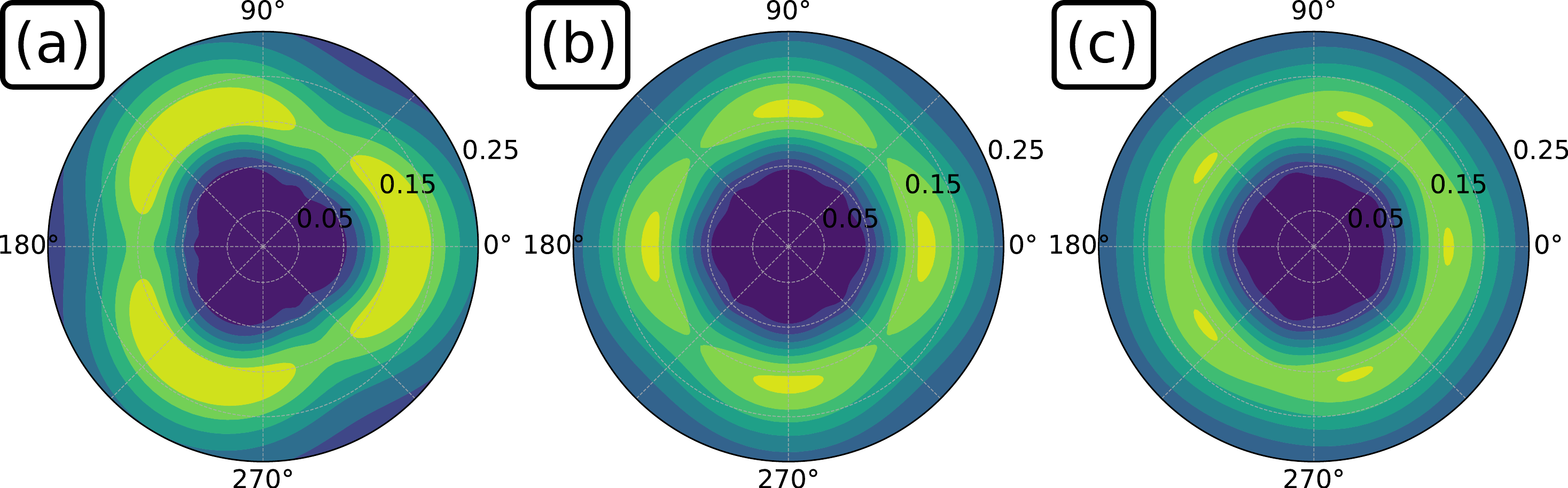}
   \caption{Specific heat (shown in colors) in terms of the charge $2\pi q$ (polar coordinates) and temperature (radial coordinate) for $N_f=3$(a), $4$(b) and $5$(c), assuming $N_c=45$. This plot was derived from ~\eqreff{eqLambdaFinal}.  }
   \label{figNf3and4and5}
 \end{figure}

Equation \eqref{eqLambdaFinal} indicates a rotation symmetry related to charge in the modular directed  network. 
These symmetries  causes the petal structures in specific heat observed in main text and in\figref{figNf3and4and5}.
 
\subsection{Origin of Asymmetries in the Specific Heat}

The results obtained in the previous section paves the way to understanding the relationship between the modular directed networks (flux) and the magnetic Laplacian spectrum, as well as the specific heat symmetry.  However, these results assume that the inner structures $\mathbf{H_{in}}$ are undirected.  The effect of directionality can be inferred by genearting random directions inside the intrablocks, i.e.~ by imposing that $[\mathbf{W_{in}}]_{u,v}$ has probability $p_c$ to take value $1$.  

\begin{figure}[!htb]
    \centering
    \includegraphics[scale=.3]{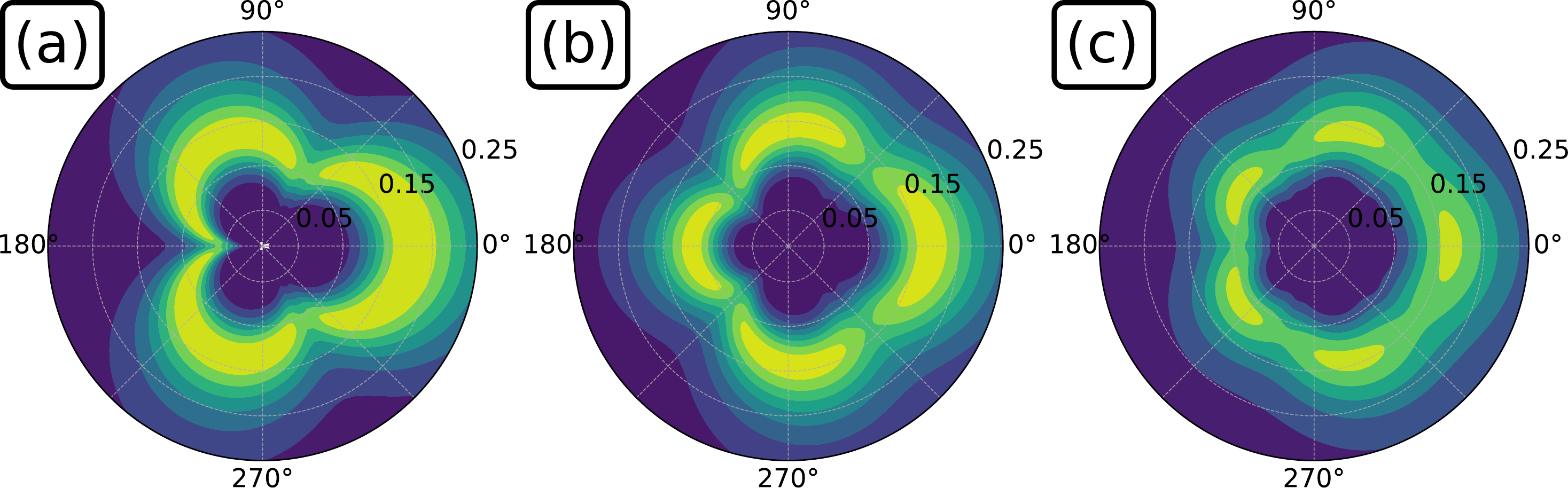}
    \caption{ Specific heat (colors) in terms of the charge 
     $2\pi q$ (angle) and temperature (radius), for $N_f=3$(a), $4$(b) and $5$(c), assuming $N_c=45$. The networks were
generated randomly, imposing the probability of having a
directed edge as $p_c=30\%$. Observe the obtained
asymmetric petals contrasting with the results obtained
previously for the uniform connections.    }
    \label{figRandomNf3and4and5}

\end{figure}

Adopting $p_c=30\%$, we calculate the specific heat by using numeric diagonalization, yielding the structures in\figref{figRandomNf3and4and5}.  We can observe that the obtained petals are not symmetric, unlike what had been observed for uniform connections.

\section{Frustration space and relationship with quantum dynamics}
\label{secSupMatEvoOperator}
In order to measure the similarity between two nodes ($u$ and $v$) we extend the analogy with quantum mechanics by defining  a parameter $t \in \mathbb{R}$, time, and the well-studied  quantum evolution operator~\cite{sakurai1995modern}
\begin{eqnarray}
\mathbf U_q(t) = \sum_{l=0}^{\infty} \frac{(-i t \mathbf{H}_q)^l}{l!}.
\end{eqnarray} 

This operator maps a given state,  $|\phi_q\rangle$, into the corresponding time evolved state $|\phi_q(t)\rangle$ through
\begin{eqnarray}
|\phi_q(t)\rangle =\mathbf U_q(t)|\phi_q\rangle.
\end{eqnarray}
Therefore,  we have a dynamical system, $(\mathbb{C}^{|V|}, U_q, \mathbb{R})$,  which can be used to characterize a given node through the state dynamics of that node at $t=0$.

We restrict our analysis to the dynamics of the phases of a given state, which is an essential entity for our dynamics because it reflects modifications of the directed edges on graph. In order to study the aforementioned dynamics, it is convenient to map the evolution operator into a smoothed version through Laplace transform, which can be 
performed by solving the integral

\begin{eqnarray}
 |\tilde \phi_q(s)\rangle =\int\limits_{0}^\infty e^{-st}|\phi_q(t)\rangle dt,
 \label{eqLaplaceTransform}
\end{eqnarray}
where $s\in\mathbb{R}$.

Using the spectral representation of the evolution operator, 
\begin{eqnarray}
\mathbf U_q(t) = \sum\limits_{l=0}^{|V|}  e^{-i \lambda_{l,q} t}|\psi_{l,q}\rangle \langle \psi_{l,q}|,
\end{eqnarray}

the~\eqreff{eqLaplaceTransform} can be expressed as 
\begin{eqnarray}
|\tilde \phi_q(s)\rangle  &=& 
    \sum\limits_{n=0}^{|V|}\left[\left(\int\limits_{0}^\infty e^{-(s +i\lambda_{n,q})t} dt \right)  |\psi_{n,q}\rangle \langle \psi_{n,q}| \right]|\psi\rangle \nonumber \\
       &=& 
    \left(\sum\limits_{n=0}^{|V|}\frac{|\psi_{n,q}\rangle \langle \psi_{n,q}|}{s+i\lambda_{n,q}} \right)  | \phi_q\rangle\nonumber\\
    &=&i\mathbf{G}_q(is) | \phi_q\rangle,
    \label{eqLaplaceEvo}
    \end{eqnarray}
   where $\mathbf{G}_q$ is known as a \emph{propagator} in the literature of many body physics and quantum transport~\cite{bruus2004many,datta1997electronic}.

\section{Relating charge and separability}\label{secSupMatChargeAndSeparability}
In this section we study the relationship between the
charge, $q$, and the separability of nodes bellowing to different communities in the $\mathbb{S}^1$ space (frustation space, see \ref{secMethodsFrustation}) embedded into $\mathbb{R}^2$. In order to do so, we create a network with $N_f=3$, $N_c=45$, $p_d=50\%$ and $p_c=40\%$. 

For a given $q$ the position of $u$th node is given by   
\begin{align}
\mathbf{r}_u(q)^T = [\cos \mathbf\theta_{q,u} \ \ \sin \theta_{q,u}],
\end{align}
where $\theta_{q, u}$ is the $u$th component of first eigenvector of associated magnetic Laplacian.

Using the formalism of scatter matrices~\cite{mclachlan2004discriminant}  we can measure the quality of the separability for a given $q$ calculating the ratio 
\begin{align}
J(q) = \frac{|\mathbf S_b(q)|}{|\mathbf S_w(q)|},
\label{eqRatioJ}
\end{align}

In~\eqreff{eqRatioJ}  the numerator is the determinant of the between-class scatter matrix which is given as
\begin{align}
\mathbf S_b(q)=\sum\limits_{i=1}^3
N_c(\boldsymbol \mu_i(q) - \boldsymbol \mu(q))(\boldsymbol \mu_i(q) - \boldsymbol \mu(q))^T,
\end{align}
where $\boldsymbol \mu(q)$ is the mean vector for all nodes.

The denominator of ~\eqreff{eqRatioJ}   is the determinant of within-class scatter matrix, given as  
\begin{align}
\mathbf S_w(q)=\sum\limits_{i=1}^3
\sum\limits_{u \in V_i}(\mathbf r_u(q) - \boldsymbol \mu_i(q))(\mathbf r_u(q) - \boldsymbol \mu_i(q))^T,
\end{align}
where $\boldsymbol \mu_i(q)$  is the mean position vector for all nodes inside the $i$-th community.

\begin{figure}[!htb]
    \centering
    \includegraphics[width=\columnwidth,keepaspectratio]{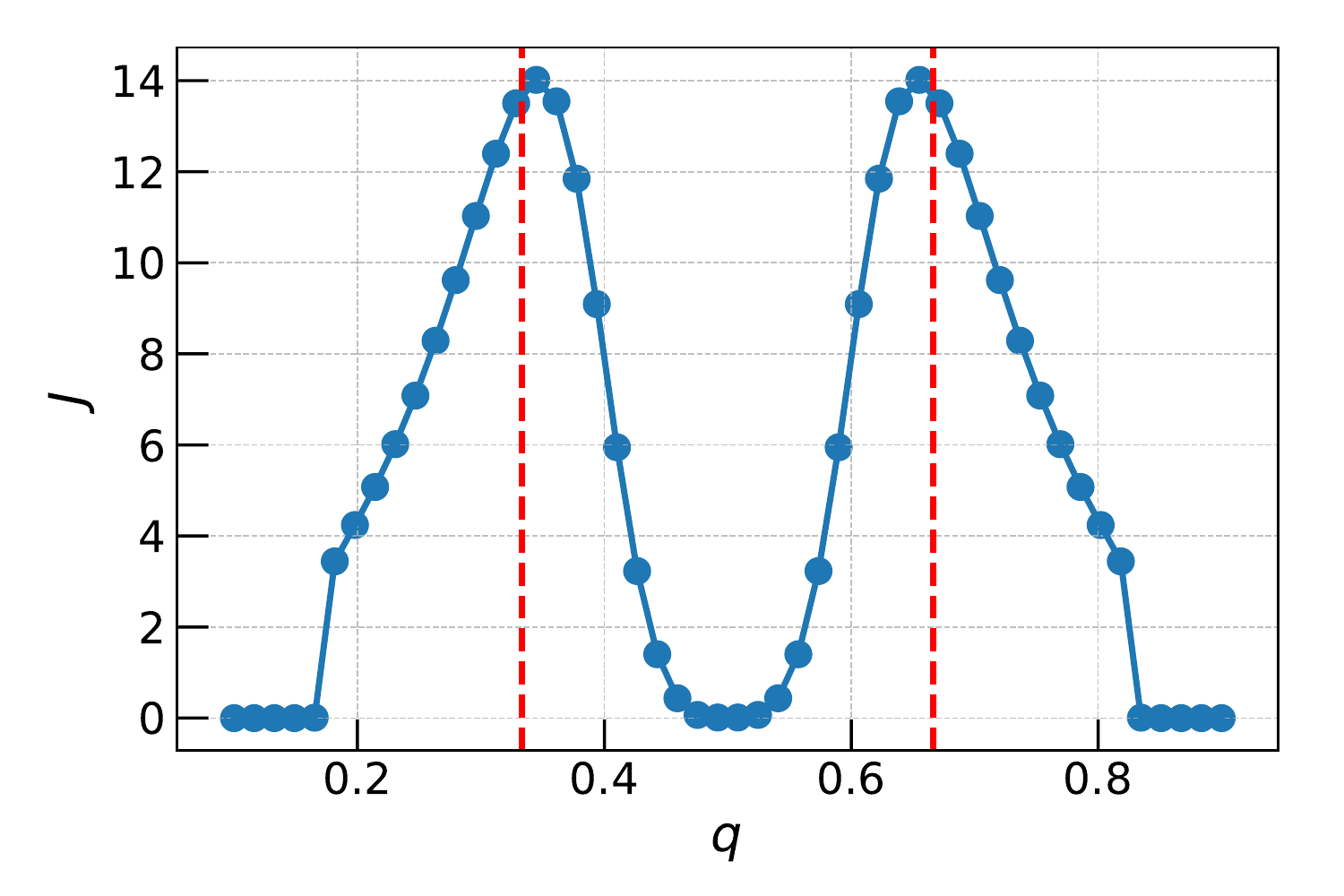}
    \caption{Quality of separability $J$ in terms of the charge value for the vertices mapped from a flux$3$ network into the frustration space, considering $N_c=45$, $p_d=50\%$ and $p_c=40\%$.  The dashed lines (in red) indicate the value of $q=1/3$ and $2/3$. }
    \label{figNf3xq}
\end{figure}

In\figref{figNf3xq} $J$ is shown in terms of $q$. Note that the best value for maximize the ratio is obtained when $q\approx \frac{1}{3}, \frac{2}{3}$ .

\section{The SOM database}
\label{secSupMatDatabase}

The database used as input to the SOM experiments includes $2000$ networks of type Watts-Strogatz (WS), Barabási-Albert (BA),  Erd\H{o}s-Réyni (ER), scale free (SF),  and modular directed networks with $3$ (flux3) and $4$ (flux4) modules.  The involved parameters are given in Table \ref{tableSupMatSom}. 

\begin{table}[!htb]
\begin{tabular}{|c|c|c|c|}
\hline
                          & \cellcolor[HTML]{EFEFEF}min & \cellcolor[HTML]{EFEFEF}max & \cellcolor[HTML]{EFEFEF}num. points \\ \hline

\cellcolor[HTML]{EFEFEF}N & 300                         & 1000                        & 700                                 \\ \hline

\cellcolor[HTML]{EFEFEF}m & 2                           & 6                           & 4                                   \\ \hline
\cellcolor[HTML]{EFEFEF}p & 4/N                         & 13/N                        & 9                                   \\ \hline
\end{tabular}
\caption{The parameters involved in the adopted models.  In the reported specific heat experiments, they are set in random fashion. $|V|$, $m$ and $p$ are, respectively, the number of vertices (or network size); the  number of outgoing edges (or out degree) for BA network and the edge probability for ER. }
\label{tableSupMatSom}
\end{table}

WS networks are generated from a regular, unidimensional network, which is rewired with probability $1\%$. After rewiring, the edges undergo a new rewiring procedure.  Now, the undirected edges are treated as two directed edges, and each of them are rewired with $10\%$ probability.  In this way, a directed version of the original network is obtained.

The specific heat for each network is calculated in the temperature range  $[0.01, 0.15]$ and charge $[0, 1/2]$, with $25$ points each.

\section{Entropic distance and permutations in  adjacency matrix}
\label{secSupMatInference}

\begin{figure}[!htb]
    \centering
    \includegraphics[width=\columnwidth,keepaspectratio]{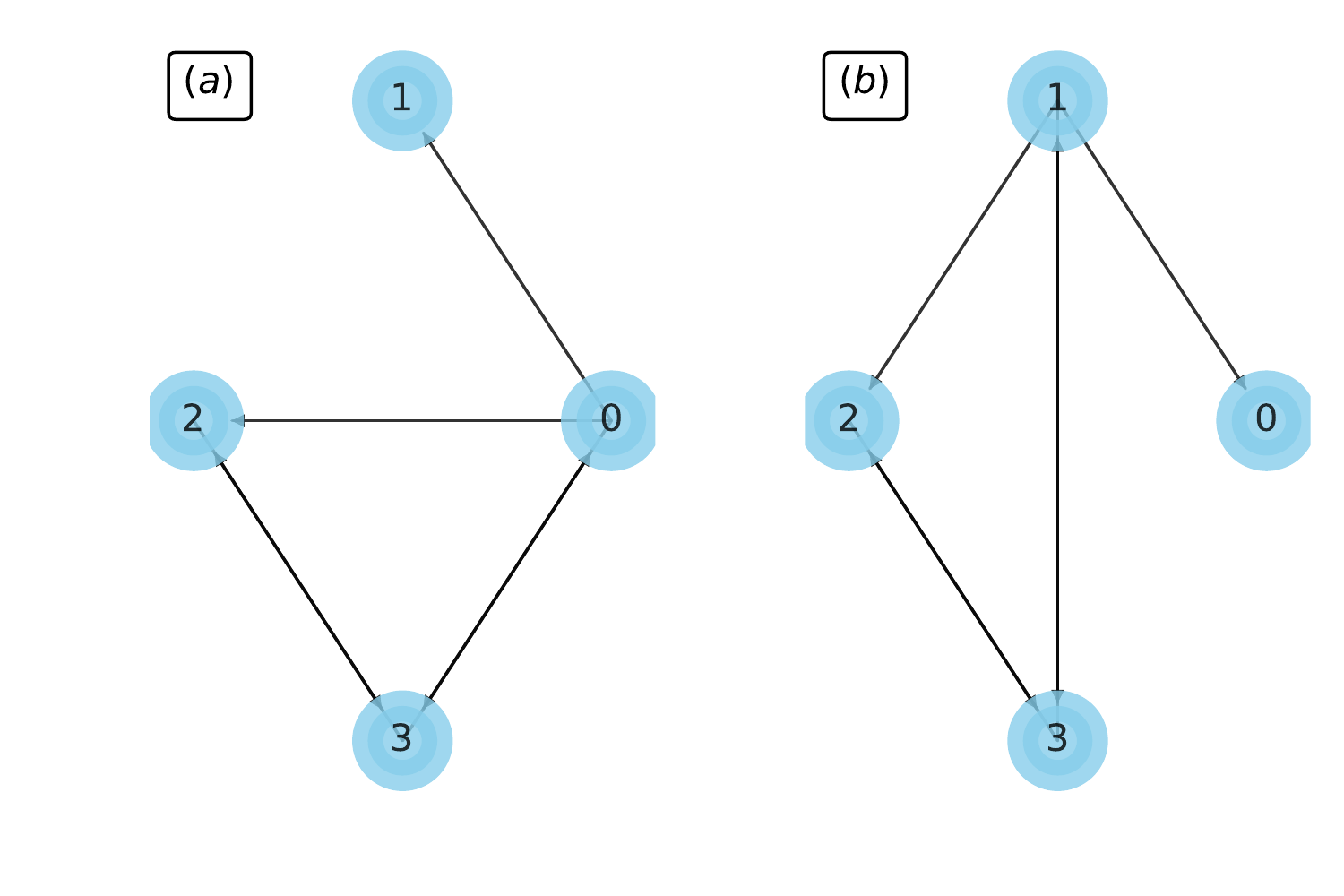}
    \caption{The graphs in (a) and (b) are isomorphic in the sense that they can be mapped one into the other by changing the indexes $0\to 1, 1\to 0$. As such, a null distance could be expected between them.}
    \label{figSameGraph}
\end{figure}

In this section we discuss network parameter estimation by using entropic distance, as in~\eqreff{eqDistEnt}.  The \figref{figSameGraph} illustrates two isomorphic graphs, $\tilde G$ (a) and $G$ (b).   The trace can be defined in the eigenvector base,~\eqreff{eqEigenVec}, associated to any of the considered graphs.  In the case of the eigenvectors
$\tilde G$, the trace operation correspondos to
\begin{align}
\mathrm{Tr}[?] = \sum\limits_{l=1}^{|V|}\langle\tilde\psi_{l,q}| ? |\tilde\psi_{l,q}\rangle .
\end{align}

Using~\eqreff{eqEnt} and trace operation, the entropy of graph $\tilde{G}$ can be expressed as
\begin{align}
S(\tilde G, q, T) = \sum\limits_{l=1}^{|V|}\frac{e^{-\frac{\tilde\lambda_{q, l}}{T}}}{\tilde Z(T,q)}\log \frac{e^{-\frac{\tilde \lambda_{q, l}}{T}}}{\tilde Z(T,q)}.
\end{align}
Having defined entropy, it is now possible to calculate the entropic distances between the two considered networks.  In order to do so, it is necessary to calculate the second right hand side of~\eqreff{eqDistEnt} .  Expanding the trace of the density matrices,~\eqreff{eqDensity}, we have

\begin{widetext}
\begin{align}
\mathrm{Tr}\left[\boldsymbol{\tilde { \rho}}_q(T) \mathrm{Log} \boldsymbol{\rho}_q(T)\right] 
&= 
\sum\limits_{l=1}^{|V|}\langle\tilde\psi_{l,q}|\left\{
\sum\limits_{m=1}^{|V|} \frac{e^{-\frac{\tilde\lambda_{q, m}}{T}}}{\tilde Z(T,q)}
|\tilde \psi_{m,q}\rangle \langle\tilde\psi_{m,q}| \mathrm{Log}\left(
\sum\limits_{n=1}^{|V|} \frac{e^{-\frac{\lambda_{q, n}}{T}}}{ Z(T,q)}
| \psi_{n,q}\rangle \langle\psi_{n,q}|
\right)
 \right\}|\tilde\psi_{l,q}\rangle  \nonumber\\
 &=
\sum\limits_{l=1}^{|V|}\frac{e^{-\frac{\tilde\lambda_{q, m}}{T}}}{\tilde Z(T,q)}
\langle\tilde\psi_{l,q}|
  \mathrm{Log}\left(
\sum\limits_{n=1}^{|V|} \frac{e^{-\frac{\lambda_{q, n}}{T}}}{ Z(T,q)}
| \psi_{n,q}\rangle \langle\psi_{n,q}|
\right)
|\tilde\psi_{l,q}\rangle 
\label{eqDistTrace}
\end{align}
\end{widetext}

The entropic distance~\eqreff{eqDistEnt}  between $\tilde G$ and $G$ yields a value $S_d \approx 2.403$ for $q=1/3$ and $T=0.5$.  However, given that the two considered graphs are isomorphic, it would be interesting to have null distance instead.  In order to define a entropic distance as a spectral distance between networks, we replace the  second right hand side of~\eqreff{eqDistEnt} by 

\begin{align}
\sum\limits_{m=1}^{|V|}\frac{e^{-\frac{\tilde\lambda_{q, m}}{T}}}{\tilde Z(T,q)}\log \frac{e^{-\frac{\lambda_{q, m}}{T}}}{ Z(T,q)}.
\end{align}

\end{document}